\documentclass[a4paper,10pt]{cas-dc}

\usepackage[authoryear]{natbib}
\usepackage{epsfig,color,amsmath}
 
\usepackage{wrapfig}
\usepackage{setspace}  
\usepackage{xcolor}
\usepackage{epstopdf}
\usepackage{amssymb}
\usepackage{url}
\usepackage{mathtools}
\usepackage{enumitem}
\usepackage{multirow}
\usepackage{makecell}
\usepackage{amsmath}
\usepackage{stackrel}
\usepackage{booktabs}
\usepackage[flushleft]{threeparttable}
\usepackage{makecell}

\usepackage{subfig}
\usepackage{graphicx}

\usepackage[linesnumbered,lined,boxed,commentsnumbered,ruled,longend]{algorithm2e}


\setlist[itemize]{leftmargin=*}

\makeatother

\newcommand{\cs}{{\cal s}}

\allowdisplaybreaks[4]
\renewcommand*{\thefootnote}{\fnsymbol{footnote}}

\usepackage{amsthm}

\usepackage[usestackEOL]{stackengine}
\stackMath
\usepackage[framemethod=TikZ]{mdframed}
\usepackage{bm}
\mdfdefinestyle{MyFrame}{%
	linecolor=black,
	outerlinewidth=1.5pt,
	roundcorner=1.5pt,
	innerrightmargin=5pt,
	innerleftmargin=5pt,}

\usepackage{tikz}
\usetikzlibrary{matrix,positioning,decorations.pathreplacing}

\usepackage[export]{adjustbox}
\usepackage{amsmath}
\usepackage{lipsum}
\usepackage{mathtools}
\usepackage{cuted}
\usepackage{physics}
\usepackage{lineno}
\usepackage{enumitem}
\def\tsc#1{\csdef{#1}{\textsc{\lowercase{#1}}\xspace}}
\tsc{WGM}
\tsc{QE}

\begin{document}
\let\WriteBookmarks\relax
\def\floatpagepagefraction{1}
\def\textpagefraction{.001}

\shorttitle{\singlespacing   {\footnotesize HydroPol2D – Distributed Hydrodynamic and Water Quality Model: Challenges and Opportunities in Poorly-Gauged Catchments}}     

\shortauthors{Gomes Jr. et. al}  


\title [mode = title]{HydroPol2D – Distributed Hydrodynamic and Water Quality Model: Challenges and Opportunities in Poorly-Gauged Catchments}  



\affiliation[1]{organization={University of São Paulo, Department of Hydraulic Engineering and Sanitation, São Carlos School of Engineering},
            addressline={Av. Trab. São Carlense, 400 - Centro}, 
            city={São Carlos},
            postcode={13566-590}, 
            state={São Paulo},
            country={Brazil}}
            
\affiliation[2]{organization={The University of Texas at San Antonio, College of Engineering and Integrated Design, School of Civil Environmental Engineering and Construction Management},
            addressline={One UTSA Circle, BSE 1.310}, 
            city={San Antonio},
            postcode={78249}, 
            state={Texas},
            country={United States of America}}   

\affiliation[3]{organization={Faculty of Engineering, Architecture and Urbanism and Geography, Federal University of Mato Grosso do Sul.},
            city={Campo Grande},
            postcode={79070–900}, 
            state={MS},
            country={Brazil}}          
            
\author[1,2]{Marcus Nóbrega {Gomes Jr.}}[
        orcid=0000-0002-8250-8195X,
        ]
\cormark[1]


\ead{marcusnobrega.engcivil@gmail.com}


\credit{Conceptualization, Methodology, Software, Validation, Formal analysis, Investigation, Data Curation, Writing - Original Draft, Writing - Review \& Editing, Visualization}

\author[1]{César Ambrogi Ferreira {do Lago}}[
            orcid=0000-0002-7841-046X,
            ]
\credit{Writing - Review \& Editing}
\ead{cesar.dolago@utsa.edu}

\author[1]{Luis Miguel Castillo Rápalo}[
        orcid=0000-0002-6241-7069,
        ]


\ead{luis.castillo@unah.hn}

\credit{Writing - Review \& Editing, Data Curation, Resources}

\author[3]{Paulo Tarso S. Oliveira}[
        orcid=0000-0003-2806-0083,
        ]


\ead{paulo.t.oliveira@ufms.br}

\credit{Writing - Review \& Editing}

\author[2]{Marcio Hofheinz Giacomoni}[
        orcid=0000-0001-7027-4128,
        ]


\ead{marcio.giacomoni@utsa.edu}

\credit{Writing - Review \& Editing, Funding acquisition, Visualization, Resources}

\author[1]{Eduardo Mario Mendiondo}[
            orcid=0000-0003-2319-2773,
            ]
\credit{Writing - Review \& Editing, Funding acquisition, Visualization, Resources}
\ead{emm@sc.usp.br}

\cortext[1]{(corresponding author)}


\begin{abstract}
Floods are one of the deadliest natural hazards and are fueled by excessive urbanization. Urban development decreases infiltration by reducing pervious areas and increases the accumulation of pollutants during dry weather. During wet weather events, there is an increase in the levels of pollution concentrations and stormwater runoff that eventually reach creeks and rivers. Polluted stormwater runoff may be sources of water supply. Modeling the quantity and quality dynamics of stormwater runoff requires a coupled hydrodynamic module capable of estimating the transport and fate of pollutants. In this paper, we evaluate the applicability of a distributed hydrodynamic model coupled with a water quality model (HydroPol2D). First, the model is compared to GSSHA and WCA2D in the V-Tilted catchment, and the limitation of the critical velocity of WCA2D is investigated. We also applied the model in a laboratory wooden board catchment, focusing on the validation of the numerical approach to simulate water quality dynamics. Then, we apply HydroPol2D in the Tijuco Preto catchment, in Sao Carlos - Brazil, and compare the modeling results with the full momentum solver of the Hydrologic Engineering Center - River System Analysis (HEC-RAS). This catchment is representative of typical small, highly urbanized, and poorly gauged catchments around the world. The implementation of the model, the governing equations, and the estimation of input data are discussed, indicating the challenges and opportunities of the application of distributed models in poorly-gauged catchments. For a 1-yr return period of rainfall and antecedent dry days and assuming an uncertainty of 40\% in the water quality parameters, the results indicate that the maximum concentration of total suspended solids (TSS), the maximum load and the mass of the pollutant washed in 30\% of the volume are, $456~\pm~260$ mg.L\textsuperscript{-1}.km\textsuperscript{-2}, $2.56 \pm 0.4$ kg.s\textsuperscript{-1}.km\textsuperscript{-2}, and $89\%~\pm~10\%$, respectively. 

\end{abstract}





\maketitle




\section{Introduction}
The spatial scale is a determinant factor to decide which tools to apply in water resources problems such as flood management, modeling, and spatial analysis of pollutants transport. Solutions to these problems typically require numerical modeling, and the quality of these models usually depends on data availability and the actual state-of-the-art conceptual models to express complex phenomena of the water cycle. 

Hydrological, hydrodynamic, and pollutant transport models are fundamental tools for decision-making about strategies focused on mitigating floods and poor water quality \citep{fan2014integraccao}. In the literature, there are a variety of models that aid in the quantification of hydrodynamic processes at different temporal and spatial scales. On the watershed scale, where these phenomena are usually expressed on larger time scales (e.g., daily), the MGB-IPH Large Basin Hydrological Model \citep{collischonn2007mgb,de2013large} is an example. At the scale of rapid response events and urban catchments, the WCA2D (Weighted Cellular Automata 2D) model \citep{guidolin2016weighted}, which uses the cellular automata approach to distribute runoff and estimate water surface flood maps, is another available model. Other fast flood models are available in the literature and focus mainly on simplifying non-linear hydrodynamic equations through assumptions such as the use of logic and linear runoff distribution rules \citep{jamali2018rapid} or by data-driven approaches such as training neural networks to predict flood inundation maps \citep{kabir2020deep}.

Process-based models are typically more laborious than rapid flood models; however, they can better model events on the urban or rural scale and are not limited to the study area where they are applied. GSSHA (Gridded Surface / Subsurface Hydrologic Analysis) \citep{downer2004gssha} and SWAT (Soil and Water Assessment Tool) \citep{arnold2012swat}, are examples of process-based models. GSSHA is often used to estimate hydrological-hydrodynamic processes and is also able to model sediment transport and fate \citep{furl2018assessment, sharif2017driven}. However, few studies have used it for water quality assessment \citep{downer2015testing}. Their approach to simulate soil detachment, sediment routing, and fines deposition is based on advection-dispersion equations, complete mixed reactors, and Shield`s law. Similarly, other models such as the Water Quality Analysis Simulation Program (WASP) also use equations based on advection-dispersion to estimate the dynamics of sediment and water quality \citep{knightes2019modeling}. 

Most of these methods require empirical parameters to represent hydraulic conditions, which can increase the complexity of the calibration due to the requirement of substantially more data, especially in poorly gauged catchments \citep{fu2019review}. Some recent examples of the application of 2-D water quantity and quality models can be found in \cite{shabani2021coupled} and \cite{yanxia2022dynamic}. The research carried out in \cite{shabani2021coupled} coupled HEC-RAS 2-D with WASP and the results illustrate a form of evaluation of the spatial distribution of soil detachment and TSS during a flood event. Using a 2-D diffusive-wave and advection-diffusion model, \cite{yanxia2022dynamic} evaluated the concentrations of total phosphorus and total nitrogen. Both aforementioned investigations, however, were feasible to be validated due to extensive available field observations of discharges, concentrations, and loads of pollutants.

In general, most studies on the dispersion and transport of pollutants address the pollution generated by agricultural sectors \citep{zia2013impact}. For instance, the SWAT model has been used to predict and analyze the impacts of agricultural management practices on the watershed scale \citep{volk2016swat}. Although able to model events on a sub-daily scale, only few articles worldwide used this model capability, and no articles with case studies in Brazil used it until 2019 \citep{brighenti2019simulating}. This shows an opportunity to investigate water quality dynamics in urban areas.

The dynamics of pollutants at urban scales is complex and requires, in addition to a complete description of physical, chemical, and biological phenomena at a proper spatial-temporal scale, a proper hydrological-hydraulic model that can explain the transport of pollutants in surface runoff. These requirements are considerable challenging in poorly gauged catchments. This could be the reason why many water quality analyzes are performed primarily with diluted metrics, such as event concentrations (EMC) or total maximum daily loads (TDML), rather than high-resolution pollutographs \citep{rossman2016storm}.

For both temporal scales (i.e., daily and sub-daily cases), a model capable of simulating water quality dynamics in a semi-distributed fashion is the Stormwater Management Model (SWMM). Although SWMM is typically applied for urban catchments, their conceptual model of semi-distributed modeling presents a limitation for the temporal-spatial distribution of pollutants in the catchment domain. Simulating hydrodynamic and water quality processes and presenting results as maps with proper resolution is  essential for understanding multiple issues. Spatial-temporal results can be used for problems such as (i) identifying prone areas to implement LIDs by estimating the potential pollutant retention, (ii) identifying areas prone to floods, and (iii) estimating pollutant concentrations in different locations in the domain. Therefore, to aid in the modeling of urbanized catchments, HydroPol2D (Hydrodynamic and Pollution 2D) is developed. The model allows hydrodynamic modeling of surface runoff and the transport of pollutants in catchments and allows estimation of water quality and quantity dynamics in user-defined temporal and spatial resolution. 

The main objective of the present work is to investigate the dynamics of surface runoff and water quality in a watershed with few available data - the Tijuco Preto catchment (TPC) in São Carlos - Brazil - and to highlight the potential of applying a distributed model in poorly gauged catchments. Total Suspened Solids (TSS) is chosen as the overall water quality indicator \citep{di2015build} and is modeled with HydroPol2D. In addition to simulating hydrodynamics and TSS transport in a poorly gauged catchment, we provide calibration and validation tests of HydroPol2D water quantity and quality components by applying the model in different case studies. To this end, the specific objectives of this paper are (i) to assess the velocity limitation of WCA2D by comparing HydroPol2D with GSSHA and WCA2D (Numerical Case Study 1), (ii) to calibrate and validate the water quality model (Numerical Case Study 2), and (iii) to compare HydroPol2D with HEC-RAS (Numerical Case Study 3). This paper is organized as follows: Sec.~\ref{sec:tc_hydro_model} presents the Material and Methods, including the mathematical description of the HydroPol2D model. Furthermore, this section presents the numerical case studies 1, 2, and 3, respectively in Sec.~\ref{sec:num_1}, Sec.~\ref{sec:num_2}, and Sec.~\ref{sec:num_3}. Results and Discussion are presented in Sec.~\ref{sec:results}, and Conclusions are presented in Sec.~\ref{sec:conclusions}.

\section{Material and Methods} ~\label{sec:tc_hydro_model}
\subsection{HydroPol2D Model}
The main concept of the model is to simulate the transport of water and mass through through the interaction between a central cell and its $4$ neighbors (Von Neumann grid) and according to their physical and morphological characteristics, it is possible to determine the variation of surface runoff along the catchment in space and time. The model consists of $3$ major sub-divisions: infiltration model (i.e., hydrologic model), non-linear reservoir + cellular automata approach (i.e., hydrodynamic model) and build-up and wash-off model (i.e., water quality model). These models will be explained in more detail in this section.

\begin{figure*}
    \centering
    \includegraphics[scale = 0.70]{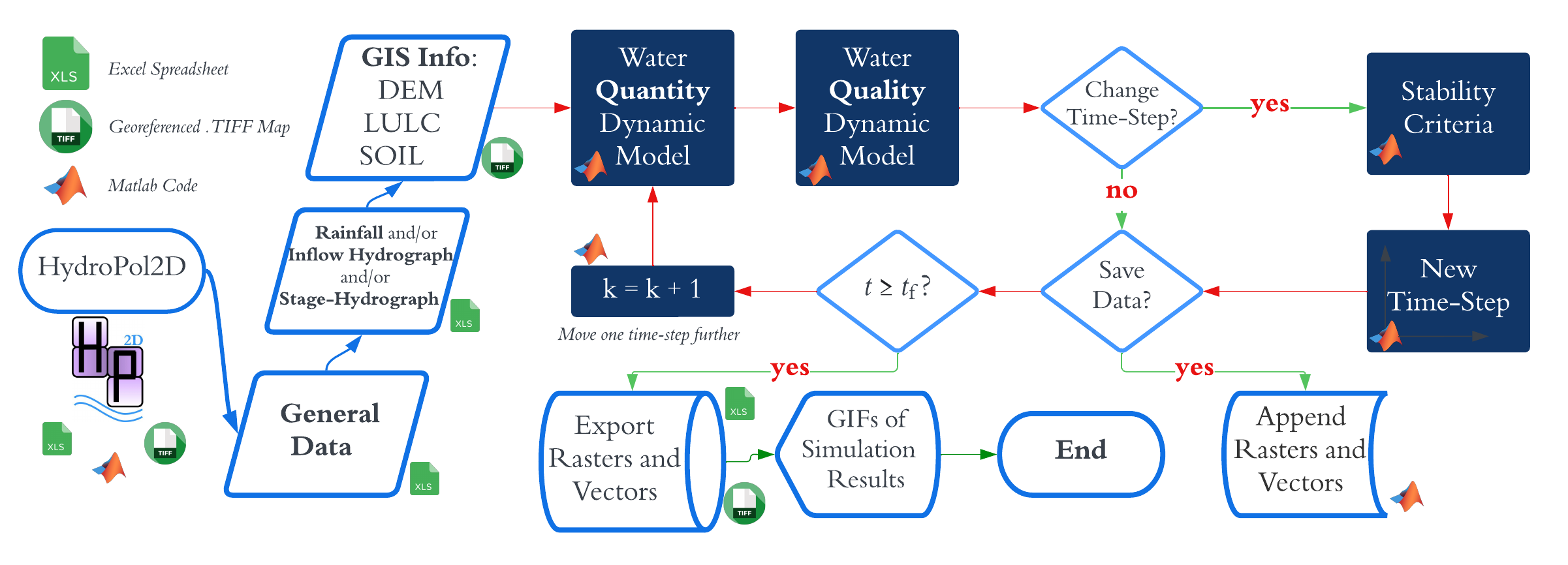}
    \caption{HydroPol2D model flowchart, where $t_f$ represents the final simulation time and $k$ represents the current time-step number. The \textit{General Data} input sets final processing parameters, stability, and all other numerical parameters, i.e., not in matrix or vector format; The \textit{Rainfall and/or Inflow Hydrograph and/or Stage-Hydrograph} sets the input rain-on-the-grid boundary conditions and/or punctual inflows and stages at internal nodes of the model. In addition, it defines the cells that receive this input hydrograph. At least one internal boundary condition has to be set. Finally, the \textit{GIS Info} input data defines the digital elevation model and the land use and occupation map.}
    \label{fig:conceptual_model}
\end{figure*}

The main parameters of the model are presented in Table~\ref{tab:inputs}. In addition, the model requires input maps that represent topography, land use and cover, and soil type. From these maps, parameters such as Manning coefficient \citep{chowv}, infiltration parameters of the Green-Ampt model \citep{green1911studies}, and water quality parameters are obtained \citep{rossman2016storm}. More details on how to obtain and estimate the parameters used in the model can be found in \citep{gomes2021spatial}, and the flowchart of the model steps is presented in Fig.~\ref{fig:conceptual_model}. First, HydroPol2D reads the input data and the boundary conditions of the rainfall and the inflow hydrograph. Then, the model discretizes the time domain and calculates two main processes: it solves the water quantity dynamic system presented in Eq.~\eqref{equ:mass_mom}, and the water quality dynamic system shown in Eq.~\eqref{equ:rate_of_change}. Following this, it decides whether to change the time step or not following Eq.~\eqref{equ:smallest_tmin}, append rasters and vectors of the main states (e.g., water depths, infiltrated depths, stored pollutant mass) and check if the simulation time ($t_f$) is reached. The numerical modeling is carried out until $t_f:= $ simulation time.

\begin{table*} 
\small
\caption{Input data typically assigned as a function of land use and occupation maps. The model requires 11 parameters to simulate the water quantity and quality dynamics} \label{tab:inputs}
\centering 
\begin{tabular}{lllll} 
\hline
\multicolumn{1}{l}{Model}                        & \multicolumn{1}{l}{Variable}     & \multicolumn{1}{l}{Symbol}                 & \multicolumn{1}{l}{Source of Uncertainty}  \\ 
\hline
\multirow{4}{*}{Hydrologic Model} & Saturated Hydraulic Conductivity & $k_{\mathrm {sat}}$ (mm . h\textsuperscript{-1}) & Spatial Variability                                                     \\ 
                                                 & Suction Head                     & $\psi $ (mm)                            & Seasonality and Soil Loss                                            \\ 
                                                 & Moisture Deficit                 & $\Delta\theta$ (cm\textsuperscript{3} . cm\textsuperscript{-3})                            & Inter-event Variability                                             \\ 
                                                 & Initial Soil Moisture            & $I$\textsubscript{0} (mm)           & Inter-event Variability                                                           \\ 

\multirow{3}{*}{Hydrodynamic Model}                    & Manning’s Roughness Coefficient  & $n$ (s . m\textsuperscript{-1/3})   & Stage Variability                                                \\ 
                                                 & Initial Abstraction              & $h$\textsubscript{0} (mm)                           & Spatial Variability                                       \\ 
                                                 & Initial Water Surface Depth      & $d$\textsubscript{0} (mm)                           & Warm-up Process                                                                  \\ 

\multirow{4}{*}{Water Quality Model}                   & Linear Build-up Coefficient      & $C$\textsubscript{1} (kg . ha\textsuperscript{-1})  & Spatial Variability                                         \\ 
                                                 & Exponential Build-up Coefficient & $C$\textsubscript{2} (day\textsuperscript{-1})      & Temporal Variability                          \\ 
                                                 & Linear Wash-off Coefficient      & $C$\textsubscript{3} (-)                            & Spatial Variability                            \\ 
                                                 & Exponential Wash-off Coefficient & $C$\textsubscript{4} (-)                            & Spatio-Temporal Variability                  \\
\hline
\end{tabular}
\end{table*}

\subsubsection{Conservation of Mass and Momentum}
The HydroPol2D model solves the mass balance equation using the diffusive wave approximation to estimate the outflow of each cell $O$ ($\mathrm{mm . h^{-1}}$) in Eq.~\eqref{equ:mass_mom}. However, the diffusive wave equation is only solved for the steepest water surface slope for each cell. Each cell can potentially have 4 flow directions and hence 4 water surface slopes. Therefore, the model solves the non-linear equation of Manning's equation (i.e., relatively computationally expensive due to power functions required) only once per cell. For the remainder of the directions, it solves the distribution of runoff through simplification using rules of cellular automata \citep{guidolin2016weighted} based on the available void volume in the boundary cells. The primary input data for the hydrological module are the spatial and temporal distribution of rainfall intensity and the input hydrograph, as well as the identification of downstream cells. Both inputs are set as external boundary conditions in the model, introducing water into the domain. Combining the main elements of the mass balance in a cell (i.e., a pixel with known resolution), we can describe the rate of change in the depth of the water in cells as \citep{rossman2010storm}:

\begin{equation} \label{equ:mass_mom}
    \frac{\partial d^{i,j}}{\partial t} = \Bigl(I^{i,j}(t) + i^{i,j}(t) - O^{i,j}(t) - f^{i,j}(t)\Bigr)\frac{1}{1000},
\end{equation}

\noindent where the indexes $i,j$ represent the position of the cell on the grid, $d^{i,j}$ is the depth of the water surface (m), $I_t^{i,j}$ is the inflow rate (mm . h\textsuperscript{-1}), $i_t^{i,j}$ is the rainfall intensity (mm . h\textsuperscript{-1}), and $f_t^{i,j}$ is the infiltration rate (mm . h\textsuperscript{-1}), given by the minimum between $C^{i,j}$ (i.e., infiltration capacity) and the inflow rate derived from the flow directions and outflow rates of each cell.

Infiltration of water into the soil is represented using the Green and Ampt (GA) model \citep{green1911studies}, which simplifies the Richards equation \citep{richards1931capillary}, and is applied to each cell of the spatial mesh created for the discretization of the catchment. More details of the GA equations and parameters are found in the Supplemental Material.

The current version of the model allows for the simulation of soil moisture restitution during dry weather periods and the spatial simulation of evapotranspiration through Pennan-Monteith simulation \citep{gomesjr2023}. Although these characteristics are not directly investigated in this article, but are available in the model repository \citep{HydroPol}. During wet weather periods, the state variable $L^{i,j}(t)$ (i.e., the saturated depth of the wetting front) is calculated only by integrating the infiltration rate over time \citep{gomes2021spatial}. Therefore, the initial value of $L^{i,j}(0)$ can be calibrated to represent the initial conditions of proper soil moisture and can be entered as an input map in the model that represents the initial conditions of soil moisture for each cell.

Both the inlet and outlet flow consider the cell topology and connections between them, following Cartesian directions in a 2D spatial mesh of Von Neumann. The conversion of depth to flow is done through the calculations of $I$ and $O$ of Eq.~\eqref{equ:mass_mom} based on the Manning equation to calculate the slope of the friction line. In this model, the friction slope is assumed to be equal to the slope of the energy line (i.e., diffusive wave \citep{vieira1983conditions}. Therefore, to distribute the volumes of surface runoff to the boundary cells, a system of weighted averages is performed in terms of the void volumes available between neighboring cells, substantially reducing the calculations by calculating the runoff velocity only for the direction of the highest slope of the water surface \citep{guidolin2016weighted} and distributing the surface runoff volume as a function of this weighted average. It is important to note that although the Manning equation is typically used for steady-state and uniform flow, it does not necessarily occur in the HydroPol2D model because the slope of the energy line is not assumed as the bottom slope. Therefore, this modeling capability allows HydroPol2D to estimate hydraulic transients.

\subsubsection{Critical Velocity Limitation}
Two versions of the HydroPol2D model were developed with respect to how the flow velocities are treated. \cite{guidolin2016weighted} restricted the velocities to critical velocities in the WCA2D model - a similar model compared to the developed model in this article. However, several studies point out that hydrodynamic modeling, especially in significant precipitation events, can present regime changes \citep{farooq2019flood}. Furthermore, two adaptations of the HydroPol2D model (a) and (b) are available and are described below:

\begin{enumerate}[]

    \item Unconstrained Velocity (HydroPol2D a) – Change of hydraulic regime is allowed; however, hydraulic jump is not modeled due to diffusive wave model that does not account for convective and local acceleration features presented in full dynamic wave models. This model assumption is more applicable for high-resolution flood inundation modeling.
    \item Constrained Velocity (HydroPol2D b) – Velocity limited to critical velocity, ensuring sub-critical fluvial regime in all cells. In this case, there are relatively lower velocities and, as a consequence, longer time-steps and shorter simulation durations. In cases where simulations are performed with relatively low spatial resolutions (e.g., $\Delta x >~\mathrm{30~m}$), the velocities in the pixels are diluted by the large pixel area and hence are typically smaller. Therefore, simulations with coarser resolutions might be typically in the sub-critical case, and HydroPol2D b) would be suitable. However, for the simulation of flood inundation maps with high resolutions, velocities can become higher than the critical velocity, and HydroPol2D a) is more appropriate.
    
\end{enumerate}

These two variations of the model result from the limitation of the maximum flow velocity, given by Eq.~\eqref{equ:maximum_flow} and Eq.~\eqref{equ:maximum_flow_2}:

\begin{equation} \label{equ:maximum_flow}
    v_m^{i,j}(t) = \min \Bigl(f_m \sqrt{g h_{ef}^{i,j}(t)}, \frac{1}{n^{i,j}}\Delta x \Bigl(h_{ef}^{i,j}(t)\Bigr)^{\frac{5}{3}}\sqrt{s_e^{i,j}(t)}\Bigr)
\end{equation}

\begin{equation} \label{equ:maximum_flow_2}
    h_{ef}^{i,j}(t) = \max \Bigl(d^{i,j}(t) - h_0^{i,j},0 \Bigr),
\end{equation}

\noindent where $v_{m}$ is the maximum velocity calculated for the steepest direction, $g$ gravity acceleration, $d$ is the water surface depth, $f_{m}$ is a factor assumed to account for models a) and b), $s_{e}$ is the slope of the water surface calculated from the water surface elevation map, $\Delta x$ is the spatial discretization of cells, $n$ is the Manning’s roughness coefficient, and $h_{ef}$ is the effective water surface depth considering losses through the initial abstraction ($h_{0}$).

In the case of model a), $f_{m}$ can be assumed to tend to infinity, such that it does not limit the flow to the critical velocity. The previous formula is applied to each time-step, for all cells of the domain, but only to the direction of the steepest water surface slope.

\subsubsection{Water Quality Modeling}
The mathematical model used to determine the transport and fate of pollutants is based on the build-up and wash-off model \citep{deletic1998first,rossman2016storm}. The term build-up refers to the accumulation of pollutants in the catchment during drought periods, and the term wash-off refers to the washing and transport of these pollutants during precipitation events or wash-off in the catchment \citep{rossman2016storm}. Several mathematical formulations for this model are proposed and, in this article, an adaptation of the exponential build-up and wash-off model is assumed. Furthermore, the increase in pollutants ($\Delta$B) in the catchment during dry weather flows is assumed as a variable dependent only on the number of consecutive dry days (ADD), as shown in Eq.~\eqref{equ:pollutant}:

\begin{equation}  \label{equ:pollutant}
    \Delta B^{i,j}_l = 10^{-4}A_c\Bigl[C_{1,l}^{i,j} \exp{-C_{2,l}^{i,j} \mathrm{ADD}}\Bigr] \pm R_l(\mathrm{ADD}),
\end{equation}

\noindent where  $C_{1}$ is the build-up coefficient, function of land use and land cover (kg . ha\textsuperscript{-1}), $C_{2}$ is the daily accumulation rate of build-up (day\textsuperscript{-1}), $\mathrm{ADD}$ is the antecedent dry days (days), $A_{c}$ is the area of (m\textsuperscript{2}), $l$ represents the classification of land use (e.g., pervious or impervious areas) and we introduce a source term $R$ to allow modeling of a non-conservative mass balance due to self-degradation or chemical reaction, varying for each land use and land cover. 

The Eq.~\ref{equ:pollutant} is valid in dry periods and calculates the build-up increment which, if added to the initial build-up, represents the amount of mass available in each cell at the end of the $\mathrm{ADD}$ time \citep{deletic1998first}. Typically, for total suspended solids, $R$ can be neglected. The original equation of the exponential wash-off model, which acts on the equation of the build-up variation during the wet weather periods, can be modeled as follows in Eq.~\eqref{equ:wash-off}

\begin{equation}   \label{equ:wash-off}
    \dv{B(t)}{t} = -W_{out}(t) = 10^{-4}A_c\Bigl(-C_3^* q(t)^{C_4^*}B(t)\Bigr),
\end{equation}

\noindent where $C_{3}^{*}$ and $C_{4}^{*}$ are wash-off coefficients in terms of specific flow rates (i.e., flow divided by catchment area) instead of flow discharges in each cell. The variable $q(t)$ is the flow rate usually given in (mm . h\textsuperscript{-1}) or (in . h\textsuperscript{-1}) and can be inferred by dividing the outlet flow by the catchment area when the catchment is modeled in a concentrated model \citep{xiao2017analytical}. The units of $C_{3}^*$ depend on the units of $q(t)$, which is used in the conversion factor of $C_{4}^*$ so that it guarantees that the wash-off rate $W$ has units of mass / time or (e.g., kg . h\textsuperscript{-1}).

The Eq.~\eqref{equ:wash-off} is used in the SWMM software and is applied in a concentrated hydrologic conceptual model, assuming a single representative value for the entire sub-catchment, as aforementioned. To represent the wash-off phenomenon, we have used a variation of the previously presented exponential model of wash-off \citep{shaw2006physical, tu2018modeling, wicke2012build, wijesiri2015influence}. The adaptation made in HydroPol2D is the following: instead of modeling the wash-off using functions dependent on specific flow rates (equivalent depth per unit of time), the model calculates the flow of pollutants, that is, loads of washing, as a function of the flow discharges leaving each cell and its available mass to be washed. Another significant difference is that pollutants enter and leave cells simultaneously during rain events. This feature changes the mass balance equation so that the equation for the rate of change of the mass of pollutants can be written as a combination of inputs and outputs of pollutant mass given by:

\begin{equation}   \label{equ:rate_of_change}
\begin{aligned}
\frac{\mathrm{\partial}B^{i,j}(t)}{\mathrm{\partial}t}  &= W^{i,j}_{\mathrm{in}}\left(t\right)\ - W^{i,j}_{\mathrm{out}}\left(t\right)  \\
&=\sum^4_{d=1}{W^{i,j}_{\mathrm{in,}\mathrm{d}}\left(t\right)}-  \sum^4_{d=1}{\overbrace{C_3{\left(Q^{i,j}_{\mathrm{d}}\left(t\right)\right)}^{C_4}f\Bigl(B\left(t\right)\Bigr)}^{W^{i,j}_{\mathrm{out,d}}\left(t\right)},}
\end{aligned}
\end{equation}

\noindent where $W$ is the wash-off load (kg . h\textsuperscript{-1}), the index \textit{in} and \textit{out} represent the inlet and outlet of the cells, respectively. The sub-index $d$ represents the flow direction, varying among leftwards, rightwards, upwards, and downwards, respectively, following the Cartesian directions. $W_{\mathrm{in,d}}(t)$ is the rate of pollutant inflow in direction $d$, $Q_{\mathrm{d}}$ is the flow discharge into this direction, and $f(B(t))$ is explained further. 

\begin{figure*}
    \centering
    \includegraphics[scale = 0.63]{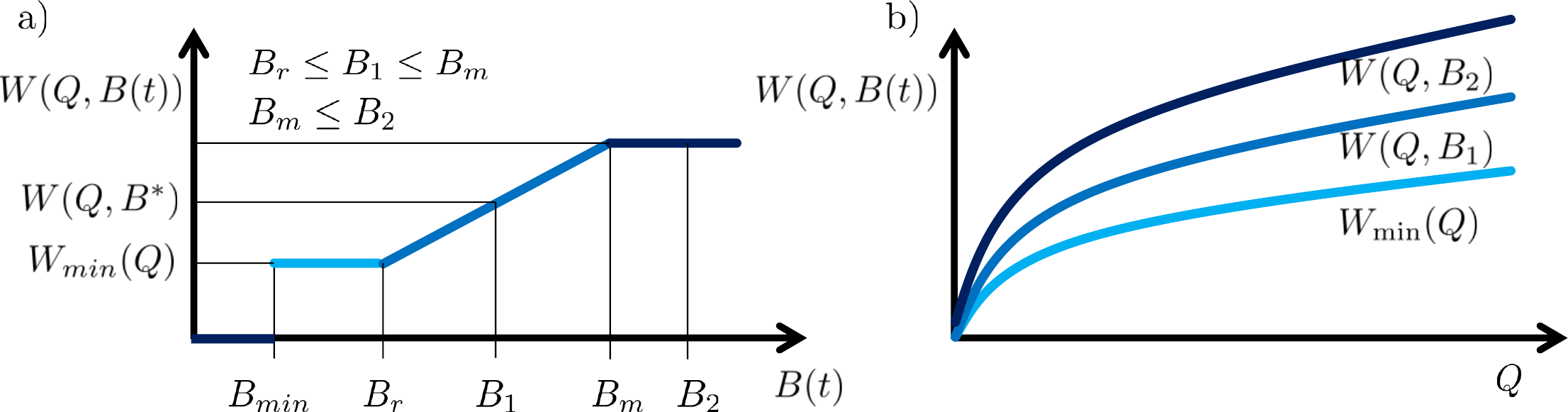}
    \caption{Scheme of pollutant washing curves. Part a) represents the washing rate as a function of accumulated mass for several cases, assuming a constant flow rate $Q$. Part b) represents the pollutant rating curve as a function of the accumulated mass in terms of the flow discharge. This figure shows the envelope of rating curves assumed for the pollutant washing.}
    \label{fig:washing}
\end{figure*}

 The function $f(B(t))$ varies the equation of pollutant washing according to the mass accumulated in the cells. For values of $B(t)$ smaller than \textit{B}\textsubscript{min}, the pollutant flux is assumed to be zero. This is the typical case of pollutants that are fixed on the soil and surface and are difficult to wash-off. For values greater than \textit{B}\textsubscript{min} but less than a threshold \textit{B}\textsubscript{r}, which depends on the type of pollutant, the washing rate follows a sediment rating curve independent of the accumulated mass; therefore, washing is exclusively dependent on the rating curve coefficients, which are equal to the wash-off coefficients. Note that $B_{\mathrm{r}}$ can be assumed equals $B_{\mathrm{min}}$, that is, the effect of the rating curve can be neglected. For the cases where the available mass is between \textit{B}\textsubscript{r} and \textit{B}\textsubscript{m}, where \textit{B}\textsubscript{m} is an upper bound, the wash rate is scaled (see Fig.~\ref{fig:washing}) by the mass of pollutants in the cell, following the typical exponential wash-off model \citep{rossman2010storm}. In the cases where $B(t)$ is greater than \textit{B}\textsubscript{m}, the maximum output rate is limited to the representative value of \textit{B}\textsubscript{m}. These conditions are expressed in Eq.~\eqref{equ:maximum_output}, such that:
 
\begin{equation}   \label{equ:maximum_output}
\small 
f\Bigl(B\left(t\right)\Bigr)=\ \left\{ \begin{array}{c}
0,\ \mathrm{if}\ B\left(t\right)\le B_{\mathrm {min}} \\ 
1,\ \mathrm{if}\ B_{\mathrm {min}}\le B\left(t\right)\le B_r \\ 
\left(1+\ B\left(t\right)-B_r\right),\ \mathrm{if}\ B_r\le B\left(t\right)\ge B_m \\ 
\left(1+B_r-B_m\right),\ if\ B\left(t\right)\ge B_r. \end{array}
\right.
\end{equation}

The imposition of a \textit{B}\textsubscript{min} value on the pollutant washing rate substantially improves the computational performance of the model by avoiding calculations in cells where the accumulated mass tends to zero and, therefore, avoiding the minimum time-step tending to zero.  Furthermore, the choice of the limit $B_r$ is effective as it ensures that pollutants follow a rating curve model for relatively low accumulated masses but larger than a minimum value \textit{B}\textsubscript{r}. If the conventional wash-off model were used (Eq.~\eqref{equ:wash-off}), when multiplying C\textsubscript{3} Q\textsuperscript{C\textsubscript{2}} by $B(t)$, with $B(t)$ tending to zero, the result would also tend to zero, which is not realistic when low pollutant mass washed by high flow, for example.

Previous modeling results indicate that for TSS, \textit{B}\textsubscript{min} $= 1$ g/m\textsuperscript{2}, \textit{B}\textsubscript{r} $= 10$ g/m\textsuperscript{2}, and \textit{B}\textsubscript{m} $= 100$ g/m\textsuperscript{2} generates consistent results. These values can also be calibrated for different pollutants. Other studies of build-up and wash-off modeling \citep{hossain2012application, wicke2012build} have applied the exponential wash-off equation presented in Eq.~\eqref{equ:wash-off} in the form of specific flow rates (i.e., outlet flow divided by the catchment area) instead of the flow discharges. However, in these studies, concentrated hydrological models of the watershed are used to represent the dynamics of surface runoff in the watershed. If we write the flow as a function of the specific flow rate ($q$), we can derive the relationship between the two modeling approaches and compare the coefficients adopted in the literature. Assuming that the specific flow rate is given in (mm . h\textsuperscript{-1}) and the modeled flow is in (m\textsuperscript{3} . s\textsuperscript{-1}), we can write Eq.~\eqref{equ:rate_flow} relating the specific flow rate to the cell outlet flow discharge, such that:

\begin{equation}   \label{equ:rate_flow}
    Q_d\left(t\right)=\left(\frac{1}{3600\times 1000}\right)q\left(t\right)\Delta x^2.
\end{equation}

Finally, analogously using Eq.~\eqref{equ:wash-off} and Eq.~\eqref{equ:rate_of_change}, we can relate the coefficients $C_3^*$ and $C_4^*$ (that is, the coefficients considering the catchment as concentrated) with $C_3$ and $C_4$ (i.e., coefficients for distributed modeling), resulting in:

\begin{equation}   \label{equ:analog_result}
    C_3=\overbrace{{\left(\frac{3600\times 1000}{\Delta x^2}\right)}^{C^*_4}}^{f_c}C_3^*,\ \ \ C_4 = C^*_4,
\end{equation}

\noindent where \textit{f}\textsubscript{c} converts C\textsubscript{3}\textsuperscript{*} developed for \textit{q}(t) in (mm . h\textsuperscript{-1}) to the model proposed here using flow discharges in (m\textsuperscript{3} . s\textsuperscript{-1}). The usual values of \textit{f}\textsubscript{c} are presented in Fig.~S1 for various values of C\textsubscript{4} and can be used for comparison between SWMM parameters and the parameters suggested in the HydroPol2D model.

The output functions of the water quality model are (i) the concentration of pollutants in time and space, (ii) the pollutant transport rate, the average concentration of the event (EMC), and the first-flush curve, that combines the normalized pollutant washed mass with the normalized runoff volume \citep{di2015build}. The governing equations for these items can be found in the supplementary material of this article.

\subsubsection{Numerical Solution and Stability}
Using the explicit and forward Euler method, that is, the information modeled at instant \textit{t} + $\Delta$\textit{t} depends only on the information at instant \textit{t}, we discretize Eq.~\eqref{equ:mass_mom} in space and time, resulting in a system of conservation equations of mass and energy resolved in two-dimensional space. For the numerical solution, either constant or adaptive time-steps can be used. The adaptive time-step values depend on the propagation conditions of the information along the cell computational mesh grid. In other words, to ensure that the information (i.e., wave propagation) does not exceed more than one cell in a time-step, the Courant–Friedrichs–Lewy (\textit{CFL}) condition is considered as the numerical stability criterion, expressed in Eq.~\eqref{equ:numerical_stability} as:

\begin{equation}   \label{equ:numerical_stability}
    \Delta t^r\left(t\right)=\mathrm{min} \left(\frac{{\alpha }^ru^{i,j}\left(t\right)}{\Delta x},\Delta t^*\right)\ \forall \ i,j\ \in \boldsymbol{\mathrm{\Pi }},
\end{equation}

\noindent where $\alpha$\textsuperscript{r} is a factor $< 1$ that ensures a Courant below the unit for the modeling of surface runoff, $\mathbf \Pi$ represents the cells domain, $\Delta$t\textsuperscript{*} is the maximum time-step assumed in the simulation, and $u$ is the wave celerity, given by Eq.~\eqref{equ:celerity}:

\begin{equation}   \label{equ:celerity}
    u^{i,j}_d\left(t\right)=v^{i,j}_d\left(t\right)\pm \sqrt{gd^{i,j}\left(t\right)},
\end{equation}

\noindent where ${v}$ is the wave velocity, and the sub-index $d$ represents an orthogonal direction. 

\begin{figure*}
    \centering
    \includegraphics[scale = 0.22]{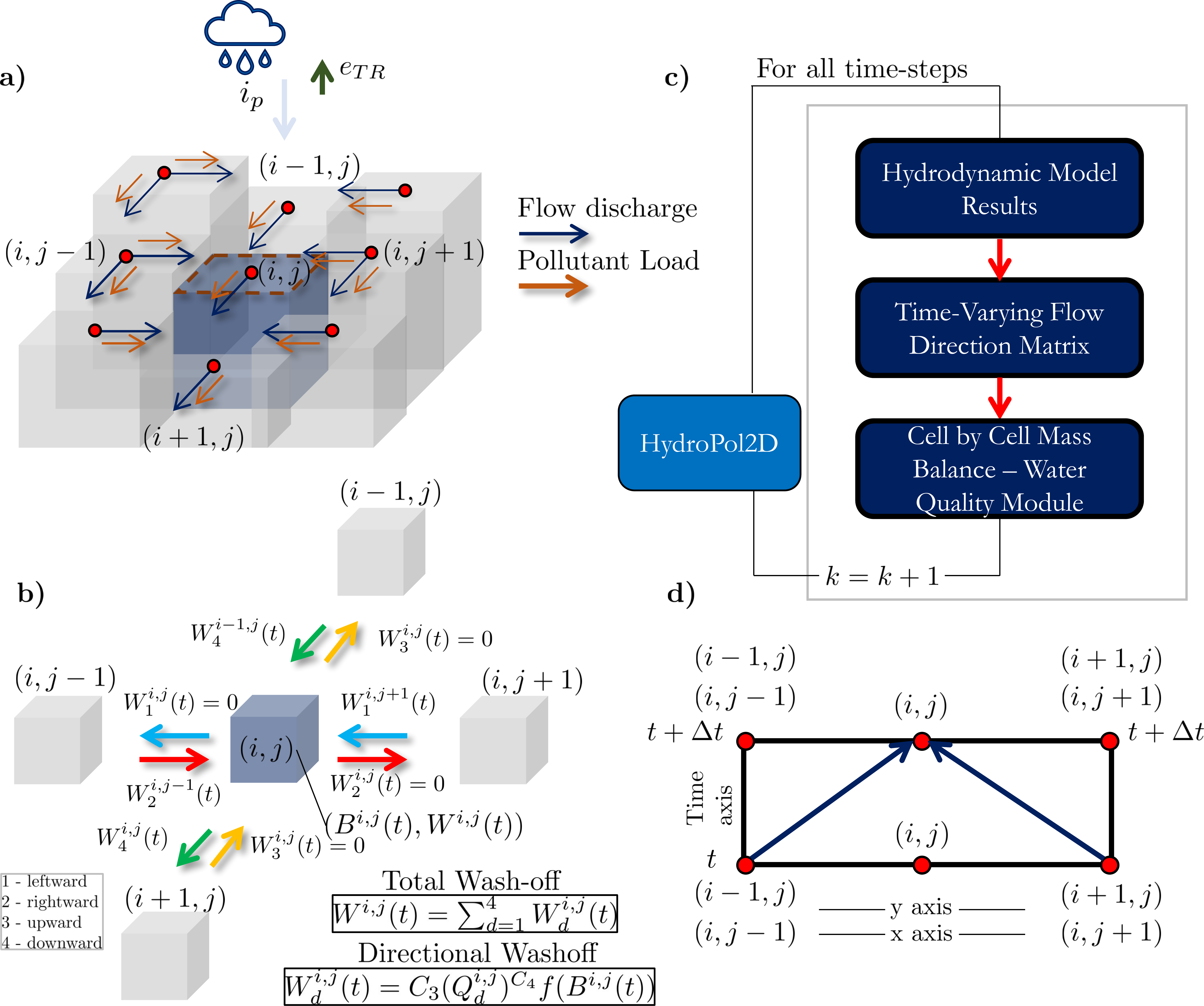}
    \caption{Scheme of the pollutant transport model, where a) represents a 3D schematic of a watershed with inflows and outflows and pollutants, b) represents a cell with pollutant outflow rates W with pollutant outflow and inflow rates as a function of the flow direction matrix, c) represents a detail of the computational scheme of the model related to water quality modeling and d) represents the computational mesh, where the water quality states of the time a posteriori depends on the states of the neighboring cells and the time to the prior time-step. Furthermore, pollutant flow rates depend on the flow rate \textit{Q}\textsubscript{d} for each direction. These flow rates are a function of the hydrodynamic model.}
    \label{fig:schematic_pollutant_transport}
\end{figure*}

Some degree of numerical diffusion occurs when using very low values of $\alpha$\textsuperscript{r} and must be previously assessed to ensure more accurate numerical solutions. For water quality, we must ensure that the available pollutant mass does not reach negative values in each time-step. This is the typical case when a long time-step is used. Fig.~\ref{fig:schematic_pollutant_transport} presents a schematic of the pollutant transport model that illustrates the processes of numerical stability and mass balances. By dividing the available pollutant mass by the pollutant wash-off for all cells in the domain, the minimum time-step is obtained to ensure numerical stability, expressed in Eq.~\eqref{equ:tmin_wq} as:

\begin{equation}   \label{equ:tmin_wq}
    \Delta t^q\left(t\right)={\mathrm{min} \left(3600\frac{{\alpha }^qB^{i,j}\left(t\right)}{{\Delta W}^{i,j}_{out}\left(t\right)},\Delta t^*\right)\ \forall \ i,j\ \in \boldsymbol{\mathrm{\Pi }}\ },
\end{equation}

\noindent where $\Delta$\textit{W}\textsubscript{out} is the outflow flux of pollutants leaving the cell (i.e., the net wash-off) considering the $4$ directions, that is, the difference between outflow and inflow of pollutant loads (kg . h\textsuperscript{-1}), and $\Delta$\textit{t}\textsuperscript{*} is the minimum time-step assumed in the model.

Theoretically, the model should not have a minimum time-step constraint $\Delta$\textit{t}\textsuperscript{*} to be considered numerically stable. However, as shown in Eq.~\eqref{equ:tmin_wq}, the time-step tends to zero as $B(t)$ approaches zero. This implies that after the first-flush, which eventually washes most of the initial pollutants out of the catchment and causes $B(t)$ to tend to zero, the time-step would also tend to zero. Therefore, we assume the minimum water quality time-step ($\Delta t^*$). Finally, the chosen time-step of the model considers the stability of both water quality and quantity models as follows:

\begin{equation}   \label{equ:smallest_tmin}
    \Delta t\left(t\right)={\mathrm{min} \left[\Delta t^r\left(t\right),\Delta t^q\left(t\right)\right]\ }.
\end{equation}

\subsection{Numerical Case Study 1 – V-Tilted Catchment} \label{sec:num_1}
\begin{figure*}
    \centering
    \includegraphics[scale = 0.29]{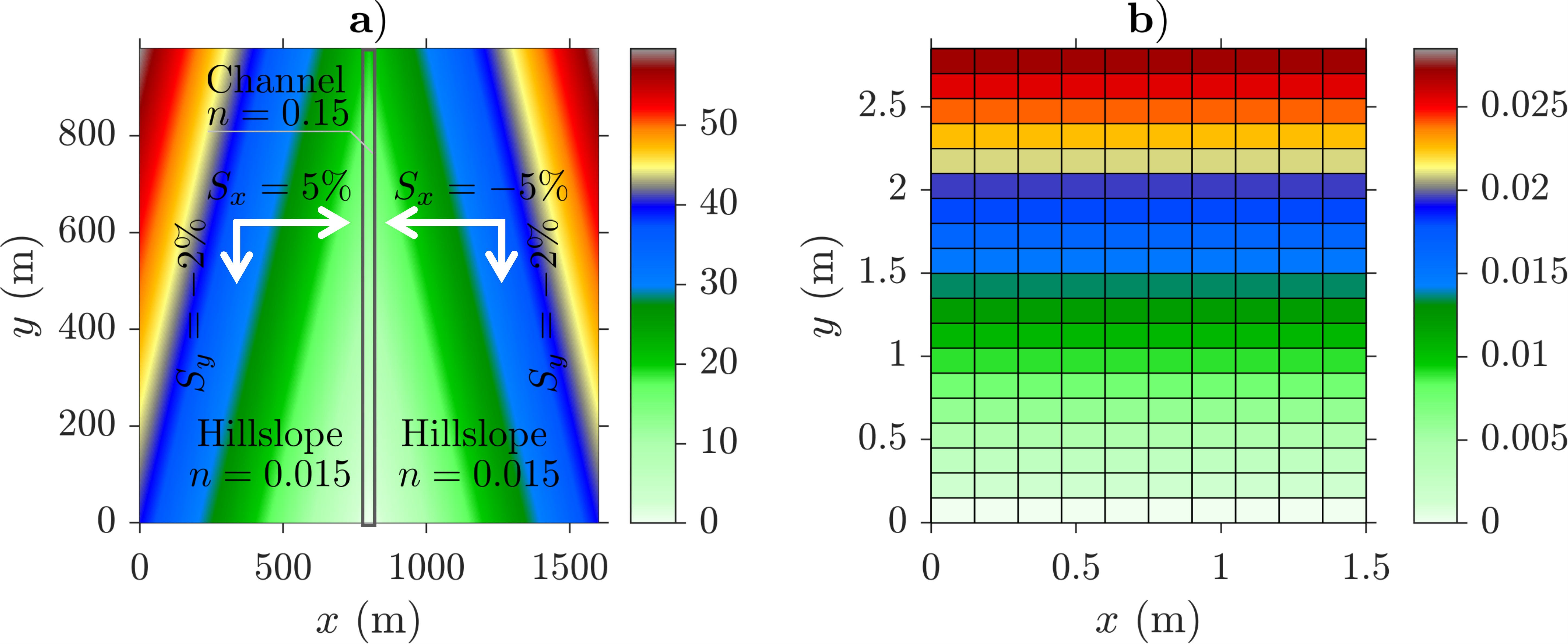}
    \caption{Catchments of Numerical Case Study 1 and 2. Part a) is the V-Tilted Catchment, with smoother hillslopes and a rougher central channel. The outlet boundary condition is assumed as normal depth with slope of 0.02. The pixel dimension is 20~m. Infiltration is not modeled and the rainfall is spatially and temporally uniform with $\mathrm{10.8~mm.h^{-1}}$ during $\mathrm{90~min}$. Part b) is the Wooden-Plane catchment \citep{zhang2020physically} with pixels of $\mathrm{0.15~m}$ with time and space invariant rainfall, and slope ($s_0$) of $1\%$, although it varies for some events assessed further. The outlet boundary condition of normal slope following the plane slope is assumed and the pixel size is 0.15~m. Infiltration is also neglected and an initial solute mass of 125 g is uniformly distributed in the catchment.}
    \label{fig:v-tilted}
\end{figure*}

The first analysis is performed in a synthetic catchment that has been used to test computer models related to surface runoff models \citep{fry2018using, junior2022flood, kollet2006integrated}. The objective of this numerical case study is twofold: assess the influence of space and time discretization and investigate the limitation of critical velocity. We tested the HydroPol2D model in the V-Tilted catchment. This theoretical catchment has only one outlet pixel and is assumed to have a width equal to the spatial discretization resolution of the cell grid ($20$ m x $20$ m). The V-Tilted catchment corresponds to a catchment of $1,620$ m x $1,000$ m (area = $1.62$ km\textsuperscript{2}) composed of two rectangular planes (i.e., hillslopes) measuring $800$ m x $1000$ m, each coupled with a vegetated channel in the connection of the two planes \citep{junior2022flood}. The slope in the $x-x$ direction is 5\%, while the slope in the $y-y$ direction is 2\%, as shown in Fig.~\ref{fig:v-tilted}a).

Two types of ground cover are assumed: channel (\textit{n} = $0.15$ s.m\textsuperscript{-1/3}) and hillslopes (\textit{n} = $0.015$ s.m\textsuperscript{-1/3}). Only surface runoff flow is evaluated; therefore, infiltration, water quality, and runoff generated by excess saturation are not modeled in this first test. A constant rainfall rate of $10.8$ mm.h\textsuperscript{-1} in $90$ minutes is applied uniformly in the catchment. The gradient boundary condition (e.g., normal flow at the outlet) was assumed for a slope equal to the natural slope of the outlet channel. The calculation time is defined as $240$ minutes to ensure the propagation of the hydrograph through the catchment. Fig.~\ref{fig:v-tilted}a) represents the digital terrain. Different time-step discretizations are tested, ranging from $0.1$ to $60$ seconds. In addition, an adaptive time-step numerical scheme is also evaluated, and simulated hydrographs with different computational meshes are compared.

\subsection{Numerical Case Study 2 - Wooden-Plane Catchment} \label{sec:num_2}
This numerical case study aims to validate the proposed distributed water quality modeling. The water quality model, however, requires a calibrated water quantity model to predict discharges, and hence the pollutant rates. We applied the HydroPol2D model in a wooden board catchment of $\mathrm{3~m}$ length and $\mathrm{1.5~m}$ width that represents an impervious surface, as shown in Fig.~\ref{fig:v-tilted}b) \citep{xiao2017analytical,zhang2020physically}. The Manning's roughness coefficient ($n$) is spatially invariant and is assumed to be equal to $\mathrm{0.04~s.m^{-1/3}}$, and the initial depth of the water is assumed to be $\mathrm{0.5~mm}$ \citep{zhang2020physically}. Rainfall is uniformly distributed in the catchment. All experimented events had a rainfall duration of $\mathrm{28~min}$. Previous modeling comparisons of HydroPol2D with flow observations in this catchment presented in \cite{xiao2017analytical} show good agreement. Therefore, the water quantity results were assumed as calibrated. In the study presented in \cite{zhang2020physically}, salt was used as the solute and all the experiments carried out were carried out evenly distributing $\mathrm{125~g}$ of salt through the wooden board. 

In this paper, we selected two cases of experiments presented in \cite{zhang2020physically} with different conditions of rainfall and slope. Events 1 to 4 have slopes of $\mathrm{0.5^{\circ}}$ with rainfall intensities of $\mathrm{24.22,~43.16,~63.81}$ and $\mathrm{76.34~mm/h}$, respectively. Events 5 to 8 have slopes of $\mathrm{2^{\circ}}$ and rainfall intensities of $\mathrm{20.76,~41.72,~78.26~,83.99~mm/h}$, respectively. Two calibration and validation tests are performed. For events 1 to 4 (i.e., $s_0 = 0.5^{\circ}$), we select event 4 for calibration and the remainder for validation. In addition, for events 5 to 8 (i.e., $s_0 = 2^{\circ}$), event 7 was used for calibration and the others for validation. To this end, we develop a calibration optimization problem minimizing the root-square mean error (RMSE) between modeled and observed salt concentrations. This procedure is fully described in the Supplemental Material. The decision variables for the optimization problem are the wash-off coefficients $C_3$ and $C_4$ and the problem is solved with the genetic algorithm for a 40 generation and population size of 100. The build-up coefficients $C_1$ and $C_2$ were not used in the calibration since the initial mass of salt is known.

\subsection{Numerical Case Study 3 – The Tijuco Preto Catchment in Sao Carlos – Sao Paulo - Brazil} \label{sec:num_3}
The third case study tested the HydroPol2D model in the Tijuco Preto catchment (TPC), in São Carlos - São Paulo. The TPC has approximately $95$\% of impervious/urbanized areas \citep{baptista2021idas}. The digital elevation model (DEM) was built based on elevation data with horizontal and vertical spatial resolution of $12.5$ and $1$ m, respectively, obtained from the Alos database Palsar \citep{rosenqvist2007alos}. The Land Use and Land Cover Map (LULC) was obtained from the mapbiomas project, available at \cite{rs12172735} and was later reclassified into two main land uses: impermeable and permeable surfaces. Subsequently, a downscaling procedure was performed, using the nearest-neighbor method, on these data from $30$ m to $12.5$ m of horizontal resolution to match the DEM spatial resolution. Despite possible errors due to data resampling, this procedure is justified because the Alos Palsar data are the product of resampling the SRTM data from $30$ to $12.5$ m. Furthermore, the delineation of flood flood inundation maps with a resolution of $12.5$ m provides a better level of detail in the modeling outputs, as it allows the capture of the flow path of streets and avenues.

This case of study is located in São Carlos - Brazil, which has experienced intense urbanization in recent decades \citep{ohnuma2014analise}. This catchment is comparable in characteristics of many highly urbanized catchments with a lack of high-resolution data on rainfall, elevation, and almost an absence of water quantity and quality observations. 

\begin{figure*}
    \centering
    \includegraphics[scale = 0.54]{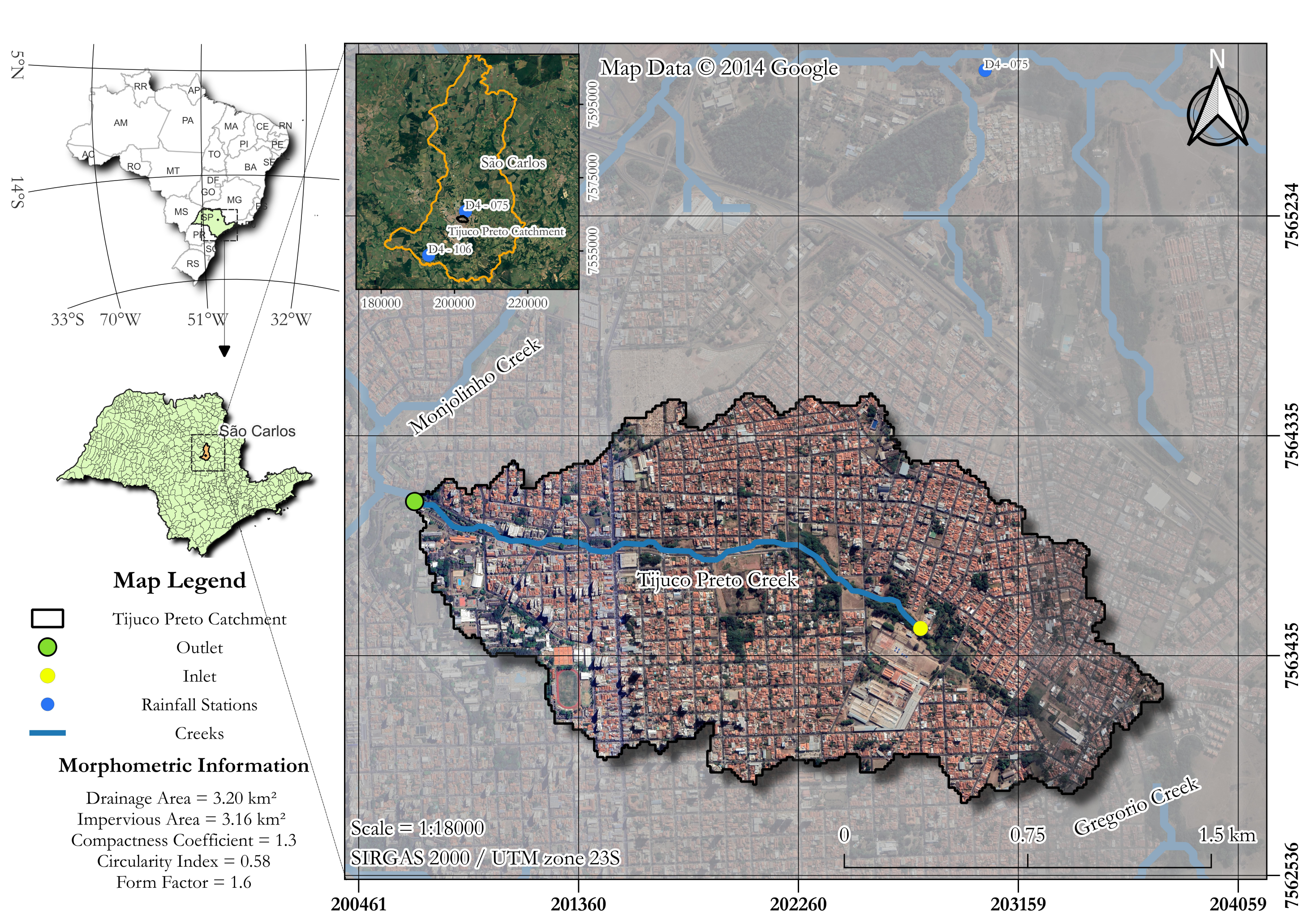}
    \caption{Tijuco Preto catchment located in São Carlos - SP. Data source: Map data © 2015 Google and IBGE.}
    \label{fig:TPC}
\end{figure*}

The modeling efforts presented here aim to explain the transport phenomenon of total suspended solids (TSS) mobilized only as a function of surface runoff. TSS was chosen due to its good representation of the general state of water quality \citep{di2015build, rossman2016storm}. To this end, the modeling of maximum water depths, maximum pollutants concentrations, and potential pollutant retention is evaluated. The TPC is shown in Fig.~\ref{fig:TPC}.

Despite the absent monitored data in this catchment, both in terms of high-resolution precipitation (e.g, sub-hourly intervals) and in terms of water depths or flows observed in the stream, the objective of this case study is to quantify in probabilistic terms the expectation of specific water depths of flood inundation depths, flow discharges, concentrations, and loads of pollutants in pixels of the catchment, especially at the outlet. The rainfall intensity boundary condition is a design storm hyetograph distributed with the alternating blocks method \citep{keifer1957synthetic}. Since the TPC is a relatively small catchment, the rainfall is modeled constant in space and variable in time.

\subsubsection{Probabilistic Distribution of Daily Rainfall and Antecedent}

For the maximum annual dry days and the subsequent creation of the ADD curve for the TPC, rainfall data was sought in the website of the Hydrological Database of the Department of Water and Electricity (DAEE), available at \cite{prodesp}. The rainfall station with prefix $\mathrm{D4-075}$ (see Fig.~\ref{fig:schematic_pollutant_transport}), named “São Carlos – SAAE”, located in the geographic coordinates $\mathrm{21^{\circ} 59'12''S, 47^{\circ}  52'33''W}$ was chosen. However, this station lacks rainfall data between $1996$ and $2013$, and, in this case, we used the station \textit{D$4-106$}, named "Fazenda Santa Bárbara" - located at coordinates $22$°$05$'$38$''S, $47$°$58$'$30$''W.

\begin{figure*}
    \centering
    \includegraphics[scale = 0.29]{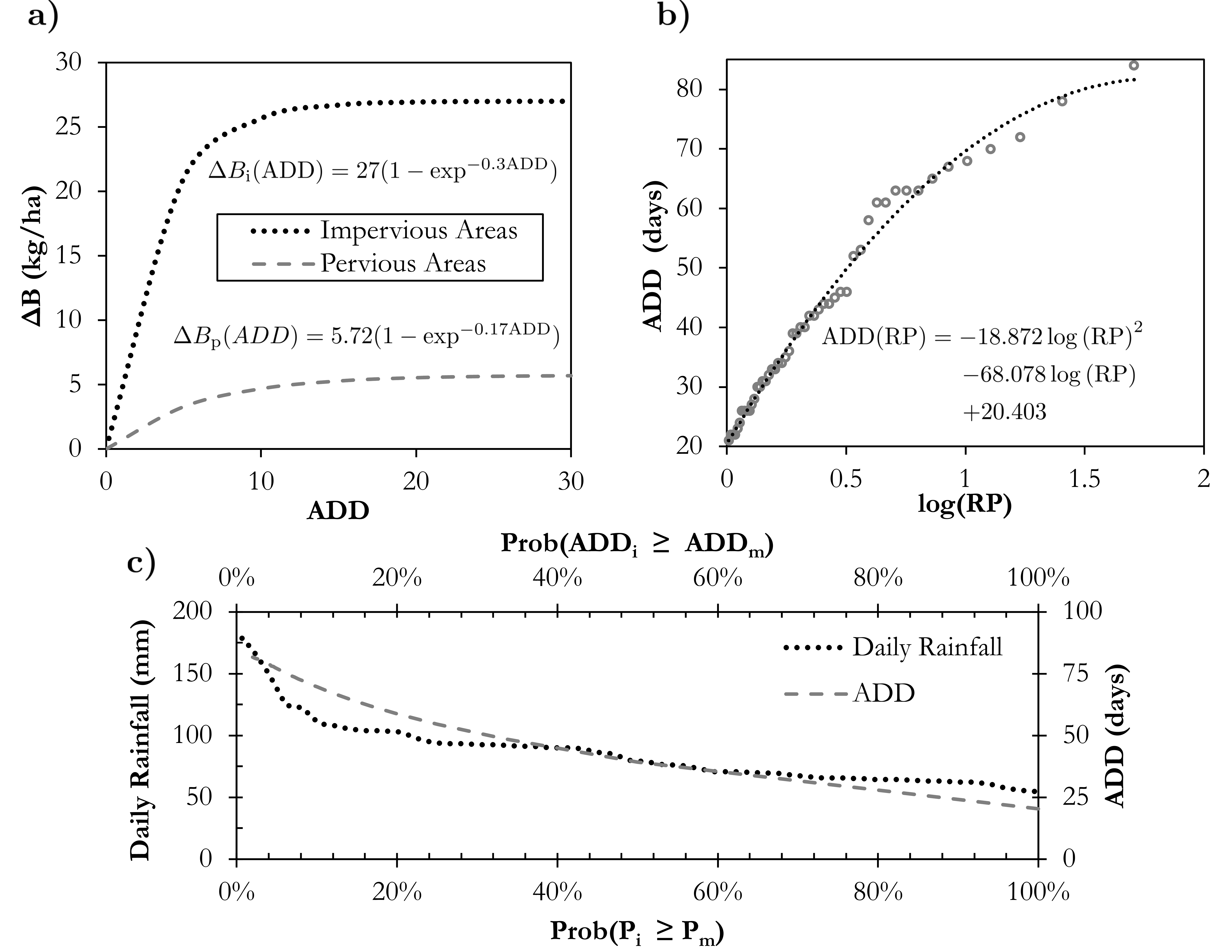}
    \caption{Accumulation of pollutants (build-up) as a function of the interval of dry days (ADD), b) Adjustment of dry days concerning empirical return periods simulated by the Weibull relation, and c) Probability distribution of ADD and daily rainfall. The rainfall is assumed space invariant in the catchment.}
    \label{fig:accumulation_pollutants}
\end{figure*}

To estimate the maximum annual dry days in the TPC, data from station \textit{D4$-075$} were used between $1970$ and $1995$, and for the years $2014$ to $2018$. For the period from $1996$ to $2013$ and $2019$, station \textit{D$4-106$} was used. Both stations do not have data for May $2016$, so this year was not used for the analysis. It is observed that the expected values of ADD are on the order of $25$ days for a $RP$ of $1$ year. The daily rainfall data presented in Fig.~\ref{fig:accumulation_pollutants} were obtained from the DAEE platform and used to fit an updated IDF curve for São Carlos \citep{gomes2021spatial}, with Sherman-type parameters of $K = 819.67$, $a = 1.388$, $b = 10.88$, and $c = 0.75$.

\subsubsection{DEM Treatment and Reconditioning}

\begin{figure*}
    \centering
    \includegraphics[scale = 0.85]{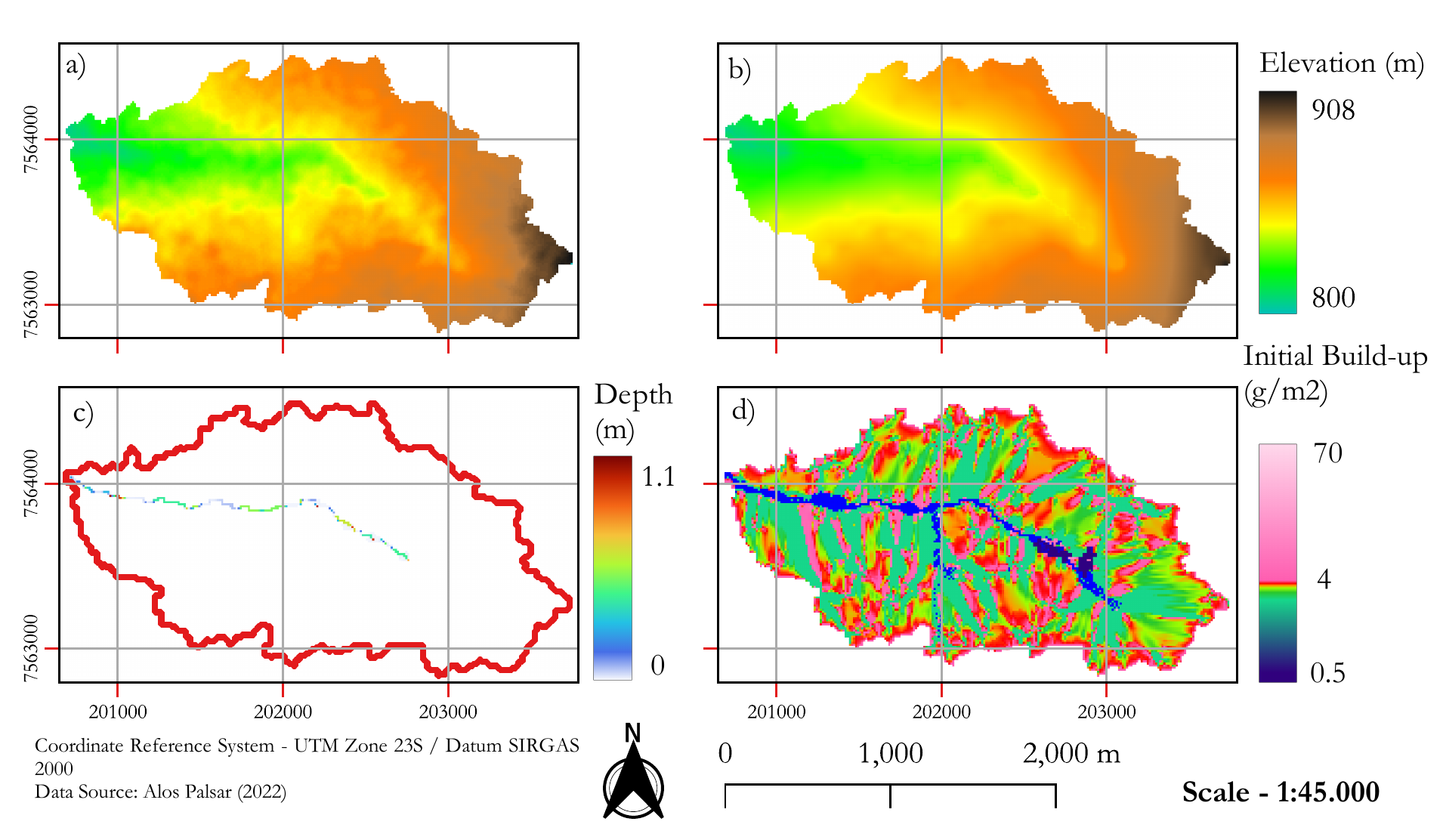}
    \caption{Relationship between the original DEM and the reconditioned DEM developed to ensure hydrological continuity and warm-up data. Part a) is the original elevation data, b) is the reconditioned DEM, c) is the warm-up depth, and d) is the initial TSS mass (initial build-up). }
    \label{fig:DEM}
\end{figure*}

Raw elevation information contains noise, accumulation points, depressions, and plateaus due to the low accuracy of the data. The elevation data was subjected to sequential processes to refine the hydraulic pathways in the catchment. First, a slope-based filter was used to remove possible noise from the elevation data, generating a raster that contains the terrain without peaks with a slope greater than $30$° (\textit{DTM filter} - SAGA \citep{passy2018use}). This slope represents an elevation difference of $7.21~m$ between the boundary cells and the cells and could represent urban features such as buildings that should be removed from the terrain model. After this operation, a raster is generated with several areas left without data, and, in the absence of such data, a bilinear interpolation filter was used to smooth the terrain lines (\textit{r.fillnuls} – GRASS \citep{lacaze2018grass}). This process ensures smoother flow lines. More details of this procedure is presented in the Supplemental Material. The final product of the procedure is shown in Fig.~\ref{fig:DEM}.

\subsubsection{Warm-Up Process and Initial Values for Modeling}
Before starting the hydrodynamic simulations, a warm-up process was simulated to represent the initial conditions of water depths in the TPS and the initial mass of pollutants in the catchment. Initial tests indicated that simulating an event with a hydrograph in the channel inlet provides better warm-up depths in the channel than a rainfall simulation on the grid (i.e., water accumulates only in the channel). Thus, a constant hydrograph with a flow of $0.3$ m\textsuperscript{3}/s for $24$ hours was simulated at the beginning of the open stream (coordinates $202762.24$; $7563794.99$ UTM $23$S shown in Fig.~\ref{fig:TPC}). This initial flow can represent an eventual base flow and clandestine sewage releases that are often released into the creek. The same inflow is also considered in rain-on-the-grid events. The downstream boundary condition of the domain was assumed to be the critical flow condition, and the outlet pixels were considered the two lowest elevation pixels on the domain boundary. The outlet represents a 25-m wide area with 2 pixels.

A different warm-up process was used to represent the initial conditions of the pollutant mass of the cells. Typically, build-up models assume that the accumulation of pollutants in the catchment is uniform for each type of land use \citep{rossman2016storm}. Therefore, in a scenario in which the entire catchment had been washed previously (e.g., a relatively large storm), for an accumulated mass equivalent to an ADD, permeable and impermeable areas would deterministically have the same accumulated mass of pollutants in each cell. A previous simulation was performed with ADD $= 10$ days and rainfall of RP $= 1/12$ years to determine more realistic conditions for the accumulation of pollutants, which is equal to the rainfall event with a probability exceedance relative to the period of $1$ month, assuming a duration of $60$ min. The hypothesis is that this event theoretically represents an initial condition of the catchment not fully washed, where a pattern of accumulation is established on the streets, buildings, and channels. Following this simulation, in each pixel of the catchment, the mass of pollutants is calculated using Eq.~\eqref{equ:pollutant}. Fig. S1, in the Supplementary Material, shows the warm-up maps for quality and quantity.

\subsubsection{Composite Design Event}
The event simulated in this study corresponds to a combination of two consecutive events: frequent annual drought (e.g., RP $= 1$ year) followed by frequent annual rainfall (e.g., RP $= 1$ year). Thus, the return period of the composite event, which corresponds to the product of each RP event, also results in RP $= 1$ year. This design event was chosen because it represents a common event in the catchment in terms of both the accumulation of pollutants and the volume of precipitation. Furthermore, more frequent rainfall events tend to produce higher average concentrations because they carry a higher amount of pollutants in a smaller volume of surface runoff \citep{di2015build}. On the other hand, very frequent events (e.g., $RP < 1/12$ years) might not even produce surface runoff to carry pollutants. The base parameters assumed in the simulation were obtained based on the literature and studies such as \cite{zaffani2012poluiccao} for the TPC, presented in Table ~\ref{tab:parameters_wq}.

\subsubsection{Parameter Estimations and Local Sensitivity Analysis}
The absence of monitoring makes the formal calibration and validation of the model intractable. The parameters of the water quantity model were assumed a priori, based on satellite information on the catchment and inspections on site. The Manning coefficient and the losses by abstraction were assumed on the basis of the land use and occupation of the catchment, classified as permeable and impermeable. Since the catchment is almost entirely impermeable (i.e., there are relatively few losses through infiltration), the calibration of the hydrodynamic model would only consider the Manning's coefficient if we neglect the effect of the initial abstraction in impermeable areas. The assumed value of the Manning's coefficient represents impermeable areas \citep{chowv}. For the water quality wash-off parameters, we perform a first estimate based on the scarce observations presented in \cite{ohnuma2014analise}. Furthermore, we evaluate the uncertainty in the wash-off parameters by a local sensitivity analysis varying the parameters $+40\%$ to $-40\%$ in terms of loads, concentrations and $\mathrm{EMC}$ of $\mathrm{TSS}$.

In addition, we compared the HydroPol2D model with the HEC-RAS full-momentum solver to check the ability of the model to predict hydrographs at the outlet. In this analysis, we simulate the same design event but without infiltration and the initial abstraction effect.

\begin{table*}
\centering
\caption{Parameters of the base scenario adopted in the simulation.}
\label{tab:parameters_wq}
\arrayrulecolor{black}
\begin{tabular}{lllllllll} 
\hline
\multirow{2}{*}{Land Use Classification} & \multicolumn{8}{c}{Parameters}                                                                                                                                                                                                                                                                                                                                                                                                                                                                                                                                                                                \\ 
\cline{2-9}
                                         & \begin{tabular}[c]{@{}l@{}}$k_{\mathrm {sat}}$ \\(mm/h)\end{tabular} & \begin{tabular}[c]{@{}l@{}}$\Delta \theta$ \\(cm \textsuperscript{3}.cm \textsuperscript{-3})\end{tabular} & \begin{tabular}[c]{@{}l@{}}$n$ \\(s.m\textsuperscript{-1/3})\end{tabular} & \begin{tabular}[c]{@{}l@{}}$h_0$ \\(mm)\end{tabular} & \begin{tabular}[c]{@{}l@{}}$C_1$ \\(kg. ha \textsuperscript{-1})\end{tabular} & \begin{tabular}[c]{@{}l@{}}$C_2$ \\(day \textsuperscript{-1})\end{tabular} & \begin{tabular}[c]{@{}l@{}}$C_3$ \\-\end{tabular} & \begin{tabular}[c]{@{}l@{}}$C_4$ \\-\end{tabular}  \\ 
\arrayrulecolor{black}\cline{1-1}\arrayrulecolor{black}\cline{2-9}
Impervious Areas                              & 0                                                              & 0                                                                                             & 0.018                                                                            & 10                                                         & 27.6                                                                                & 0.2                                                                              & 1200                                                    & 1.2                                                      \\
Pervious Areas                                & 10                                                             & 0.4                                                                                           & 0.100                                                                            & 20                                                         & 5.72                                                                                & 0.17                                                                             & 1200                                                    & 1.2                                                      \\
\hline
\end{tabular}
\arrayrulecolor{black}
\end{table*}

\section{Results and Discussions} \label{sec:results}
\subsection{Numerical Case Study 1: The Role of Velocity Limitation and Numerical Stability}
\begin{figure*}
    \centering
    \includegraphics[scale = 0.225]{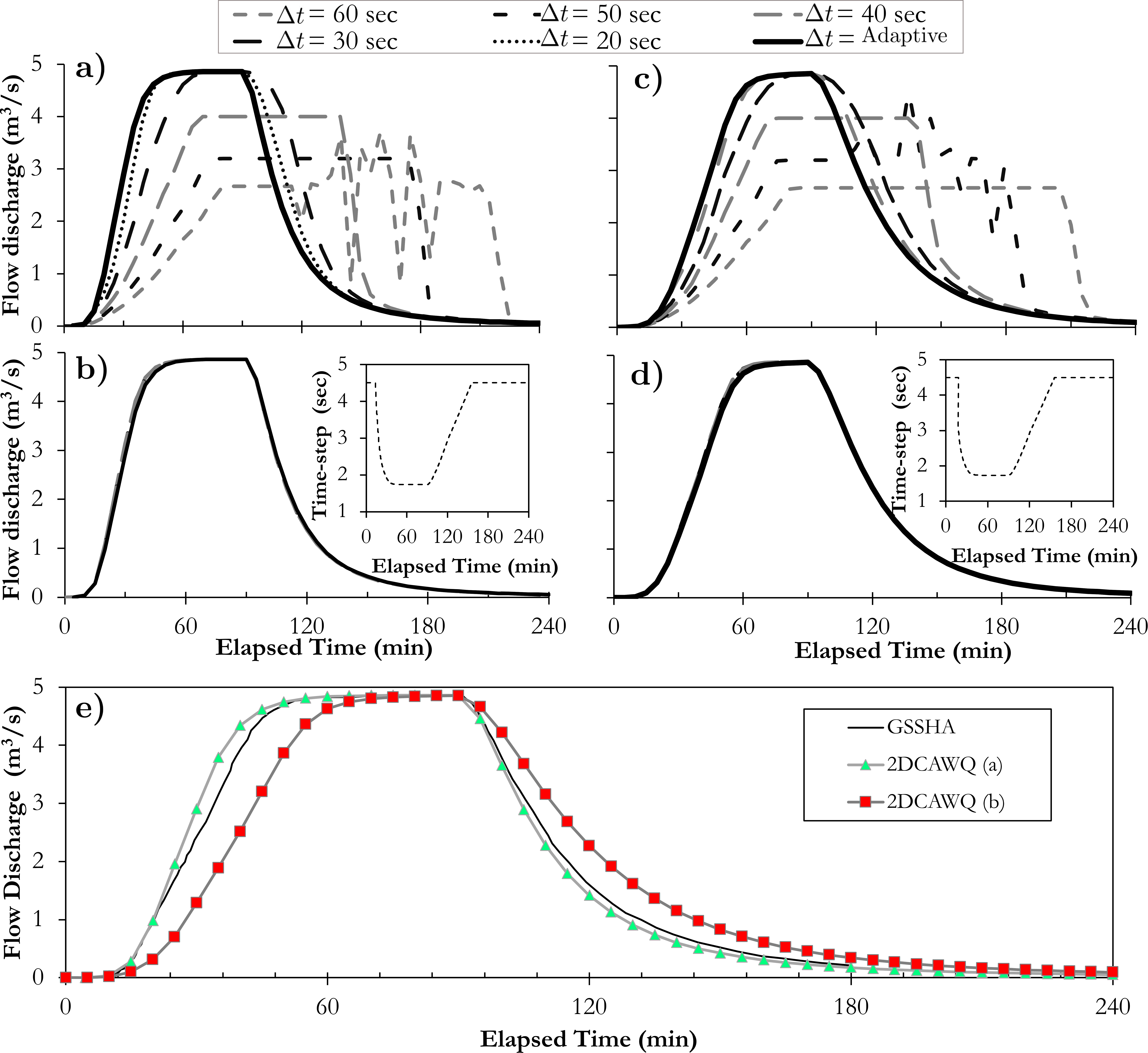}
    \caption{Comparison of hydrographs generated by different computational meshes; where (a)-(b) represent hydrographs for simulated meshes in HydroPol2D a), (c)-(d) represent simulated hydrographs for model HydroPol2D b), and (e) shows the comparison of both models, with adaptive time-step, concerning the results obtained in the GSSHA software.}
    \label{fig:hydrographs}
\end{figure*}

The mathematical model is developed by numerical discretization of differential equations solved by explicit finite differences in a forward Euler fashion. Thus, this section aims to assess the impact of different temporal discretizations on the hydrodynamic modeling of the V-Tilted Catchment, typically used to assess the performance of hydrologic and hydrodynamic models. In this analysis, several time steps were used to evaluate the numerical validation of the solution considered, limiting or not limiting the velocity to the critical velocity. Since we use forward Euler's discretization method, care must be taken to select the proper computational temporal meshgrid because the method is unconditionally unstable. In this section, we compare several hydrographs with constant time-step, with guaranteed stability and evident instability, with simulations made using the adaptive stable time-step scheme, as illustrated in Fig.~\ref{fig:hydrographs}.

The different computational meshes used in the model reveal that the surface runoff modeling is practically invariant to cases where stable time-steps are chosen (see parts b and c in Fig.~\ref{fig:hydrographs}). This implies that once guaranteed the \textit{CFL} conditions; the model can accurately predict hydrographs at the catchment outlet. The same is not true when we choose time-steps greater than $20$ seconds. The system starts to show divergences from this value for both HydroPol2D models (a) and (b), generating a total loss of accuracy and numerical instability of the method for a time-step of $1$ minute.

Significant differences occur when the HydroPol2D model restricts its maximum wave velocity. Although theoretically considering the hydraulic regime change would mean relatively smaller velocities and, therefore, allow longer time-steps, it is not consistent with the reality of more intense flow phenomena, especially in the case of large floods with high velocities. In these cases, the modeling allowing regime switching is closer to the results simulated with the GSSHA, assumed as the base scenario in this case study. Both HydroPol2D models a) and b) accurately predicted the peak flow; however, only model (a) was able to capture the time to peak more accurately as it did not limit the flow velocity. The HydroPol2D model (b) is identical to the model proposed by \cite{guidolin2016weighted}, except that the HydroPol2D model allows one to calculate infiltration, water quality, and simulate different uses and land cover.

\subsection{Numerical Case Study 2: Water Quality Model Validation}
\begin{figure*}
    \centering
    \includegraphics[scale = .71]{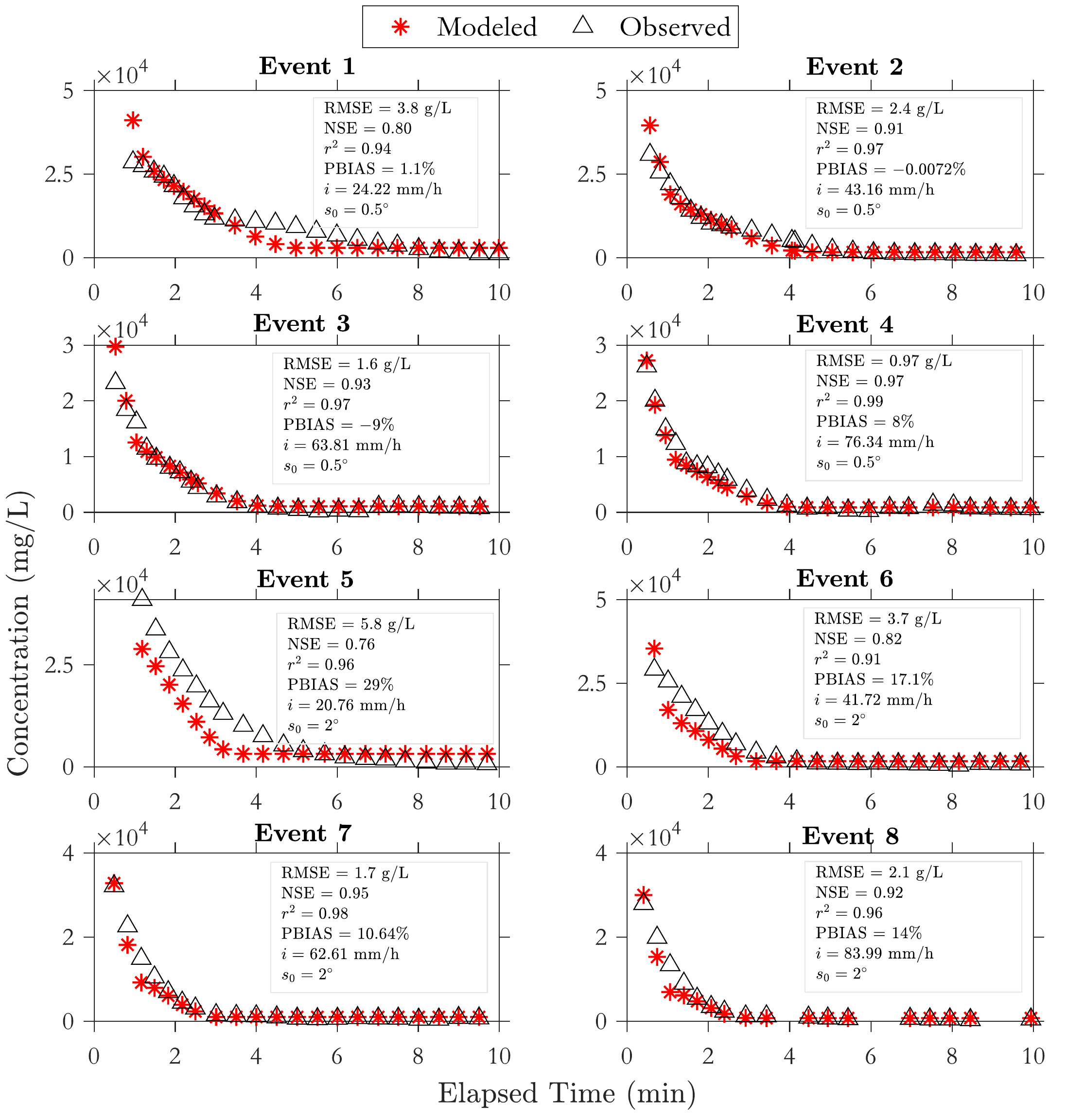}
    \caption{Comparison between HydroPol2D pollutographs with observed concentration of salt in a laboratory wooden board catchment of $\mathrm{4.5~m^2}$ \citep{zhang2020physically}. The initial mass of salt is $\mathrm{125~g}$ and is uniformly distributed.}
    \label{fig:pollutograph}
\end{figure*}

The results of the numerical calibration are presented in detail in the Supplemental Material. The pollutographs of all eight events simulated with the statistics of the root mean square error (RMSE), Nash-Suctcliffe-Efficiency (NSE) \citep{nash1970river}, coefficient of determination ($r^2$) \citep{nagelkerke1991note}, and PBIAS  \citep{gupta1999status} are presented in Fig.~\ref{fig:pollutograph}. The temporal dynamics of the solute was properly captured by the HydroPol2D model. The resulting calibrated parameters for events 1-4 are $C_3 = 9036.83$, while for events 5-8, $C_3 = 7445.11$ and $C_4 = 0.1916$. Although HydroPol2D can accurately capture the dynamics of the solute, calibration of water quality parameters is required and varies according to the physiographic characteristics of the catchment, such as slope, length, width, and roughness \citep{xiao2017analytical,zhang2020physically}.

\subsection{Numerical Case Study 3: Dynamics of Water Quantity and Quality in Poorly Gauged Catchments}

\subsubsection{Comparison Between HEC-RAS and HydroPol2D}
\begin{figure}
    \centering
    \includegraphics[scale = 0.17]{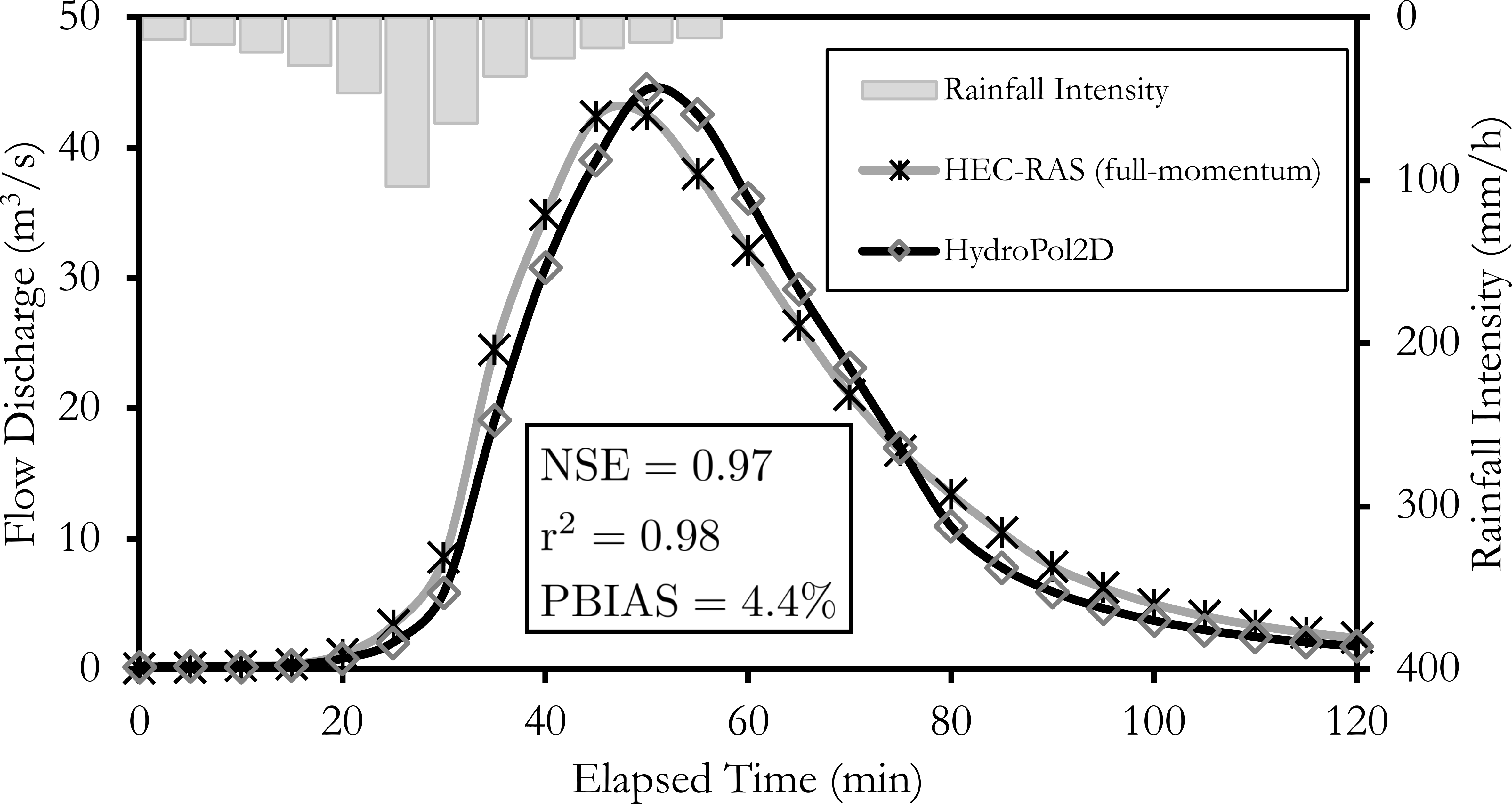}
    \caption{Outlet hydrograph comparison between the full momentum solver used in HEC-RAS and the diffusive-like numerical solution approach used in the HydroPol2D model. Both outlet boundary conditions were assumed as normal depth with gradient slope of 2\% and the catchment hydrological processes were simulated without infiltration and initial abstraction.}
    \label{fig:HECRAS}
\end{figure}

The results indicated in Fig.~\ref{fig:HECRAS} show the goodness of fitness between the full momentum solver HEC-RAS applied in the Tijuco Preto catchment compared to the HydroPol2D model. The NSE index is 0.97, the $r^2$ is 0.98 and the PBIAS is 4.4\%, indicating a good agreement between both models for all evaluated metrics.

\subsubsection{Local Sensitivity Analysis}
\begin{figure*} 
    \centering
    \includegraphics[scale = 0.66]{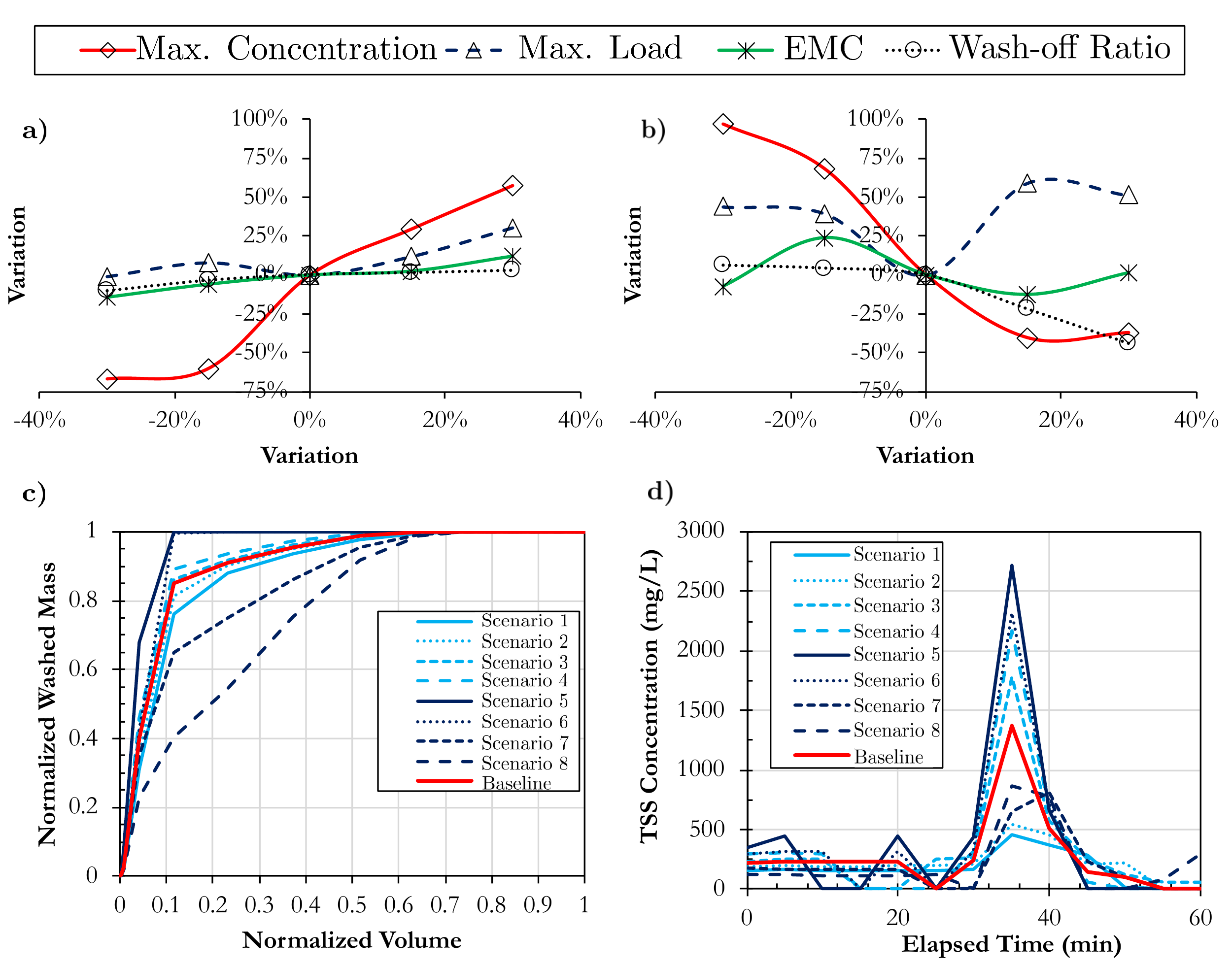}
    \caption{Results of the model sensitivity analysis for an event of RP = $1$ year, C\textsubscript{3} = $1,200$ and C\textsubscript{4} $= 1.2$. Parts a) and b) represent the sensitivity of the maximum concentration, maximum load, and average concentration of the event for variations of C\textsubscript{3} (a) and C\textsubscript{4} (b). Parts c) and d) represent the first-flush curves and pollutographs of each simulated scenario, detailed in Table ~\ref{tab:sensitivity}. The wash-off ratio is defined as the ratio between the washed mass and the available mass in the catchment.}
    \label{fig:sensitivity}
\end{figure*}

Although the parameter estimates are based on previous studies \citep{ohnuma2014analise}, a local sensitivity analysis was carried out to identify the most sensitive parameters in the water quality model. Fig.~\ref{fig:sensitivity} a) shows the sensitivity of C\textsubscript{3}, which was more sensitive to changes in maximum concentration. However, the results of Fig.~\ref{fig:sensitivity} a) indicate that the wash-off coefficient (i.e., the ratio between the washed mass and the initially available mass) was not very sensitive to this variation, suggesting that the error in this parameter does not have a large effect on the total washed mass at the outlet. Both the $\mathrm{EMC}$ and the maximum load had a low sensitivity to C\textsubscript{3}, indicating that its error does not compromise the average and diluted analyzes (e.g., EMC), but the dynamic ones such as the maximum concentration.

The results presented in Fig.~\ref{fig:sensitivity} b) show an opposite scenario than that shown in Fig.~\ref{fig:sensitivity} a). However, in general, decreasing C\textsubscript{4} increases peak concentrations and loads, which is explained by a greater mass swept at flow rates smaller than $1$ m\textsuperscript{3}/s (see Eq.~\ref{equ:rate_of_change}). Since the wash-off is a flow-dependent rating curve, lower C\textsubscript{4} exponents at flows lower than unity (i.e., $1$ m\textsuperscript{3}/s) increase the washed rates. Thus, larger masses washed in smaller volumes tend to increase the concentration. This is a numerical characteristic of the wash-off model used in this article. Another mathematical alternative to pollutants that do not follow the proposed rating curve is to add a factor $\mu$ to the flow, so that the flow used in the modeling of pollutants is ($Q + \mu$), in order to avoid this numerical problem.

The maximum load rates of $\mathrm{TSS}$ increase with increasing $C_4$, indicating a higher instantaneous washing rate at the outlet in a given time. However, these loads occur only at large flows greater than unity; therefore, the increase in $C_4$, despite increasing the maximum load, decreases the wash-off coefficient because most flows are smaller than $1$ m\textsuperscript{3}/s. This implies that higher values of $C_4$ work well on heavy pollutants mobilized in large flows; however, as pollutants are mobilized only in large flows, the total mass washed is less than a case of lower $C_4$. On the basis of this same hypothesis, $C_4$ is concluded to have a strong relationship with the density and mean diameter of the pollutant.

Fig.~\ref{fig:sensitivity} c) presents the first-flush curve for each scenario evaluated. The critical cases of the first flush (that is, larger masses washed in smaller volumes) are more evident in the variation of C\textsubscript{3} (scenarios $1$, $2$, $3$). In all cases except for scenario $8$, more than $60$\% of the pollutants were washed with $30$\% by volume \citep{di2015build}, indicating a strong first-flush. This implies that even with eventual changes in the wash-off parameters, the first flush effect is mostly observed as a result of the high impervious rate of the catchment, which quickly washes the pollutant toward the outlet. The pollutograph showed high variability, as shown in Fig.~\ref{fig:sensitivity} d), with higher peaks for higher values of C\textsubscript{4} and C\textsubscript{3}. 

Therefore, if we consider a maximum uncertainty of $40$\% in the water quality parameters, Fig.~\ref{fig:sensitivity}  (c and d) would represent first-flush and pollutogram envelopes for the simulated event. Statistically, this indicates that in $30$\% of the volume, $89$\% ± $10$\% of the $\mathrm{TSS}$ of the catchment is swept away. Similarly, the maximum load and the maximum concentration of TSS are $8.22$ ± $1.29$ kg.s\textsuperscript{-1} and $1,460$ ± $832$ mg.L\textsuperscript{-1}, respectively, and the wash-off coefficient and $\mathrm{EMC}$ are $0.63$ ± $0.11$ and $131.59$ ± $16$ mg.L\textsuperscript{-1}, respectively. Normalizing these values by the catchment area, the Load $= 2.56$ ± $0.4$ kg.s\textsuperscript{-1}.km\textsuperscript{-2}, TSS $= 456$ ± $260$ mg.L\textsuperscript{-1}.km\textsuperscript{-2}, $\mathrm{EMC}$ = $41$ ± $5$ mg.L \textsuperscript{-1}.km\textsuperscript{-2}. These values are within the expected values for moderate rainfall in urbanized areas \citep{rossman2016storm}.

\begin{table*} 
\centering 
\caption{Data used in the sensitivity analysis and its respective modeling results.}
\label{tab:sensitivity} 
\arrayrulecolor{black}
\begin{tabular}{ccccccc} 
\hline
Scenario & $\textit{C}_3$ & $\textit{C}_4$ & \begin{tabular}[c]{@{}c@{}}Maximum instantaneous\\~Concentration (mg/L)\end{tabular} & \begin{tabular}[c]{@{}c@{}}Maximum Load\\~(kg . h \textsuperscript{-1})\end{tabular} & \begin{tabular}[c]{@{}c@{}}EMC\\~(mg/L)\end{tabular} & Washoff-Ratio  \\ 
\hline
1        & 840         & 1.2         & 455.19                                                                               & 6.25                                                                                 & 112.55                                               & 0.62           \\
2        & 1020        & 1.2         & 548.33                                                                               & 6.81                                                                                 & 123.83                                               & 0.67           \\
3        & 1380        & 1.2         & 1786.73                                                                              & 7.06                                                                                 & 135.18                                               & 0.70           \\
4        & 1560        & 1.2         & 2174.31                                                                              & 8.22                                                                                 & 148.11                                               & 0.71           \\
5        & 1200        & 0.84        & 2716.02                                                                              & 9.09                                                                                 & 121.85                                               & 0.74           \\
6        & 1200        & 1.02        & 2319.25                                                                              & 8.81                                                                                 & 163.13                                               & 0.72           \\
7        & 1200        & 1.38        & 817.91                                                                               & 10.04                                                                                & 114.81                                               & 0.54           \\
8        & 1200        & 1.56        & 862.95                                                                               & 9.55                                                                                 & 133.29                                               & 0.39           \\
Baseline    & 1200        & 1.2         & 1377.75                                                                              & 6.31                                                                                 & 131.81                                               & 0.69           \\
\hline
\end{tabular}
\arrayrulecolor{black}
\end{table*}

\subsubsection{Simulation Results for RP = 1 year}
\begin{figure*}
    \centering
    \includegraphics[scale = 0.85]{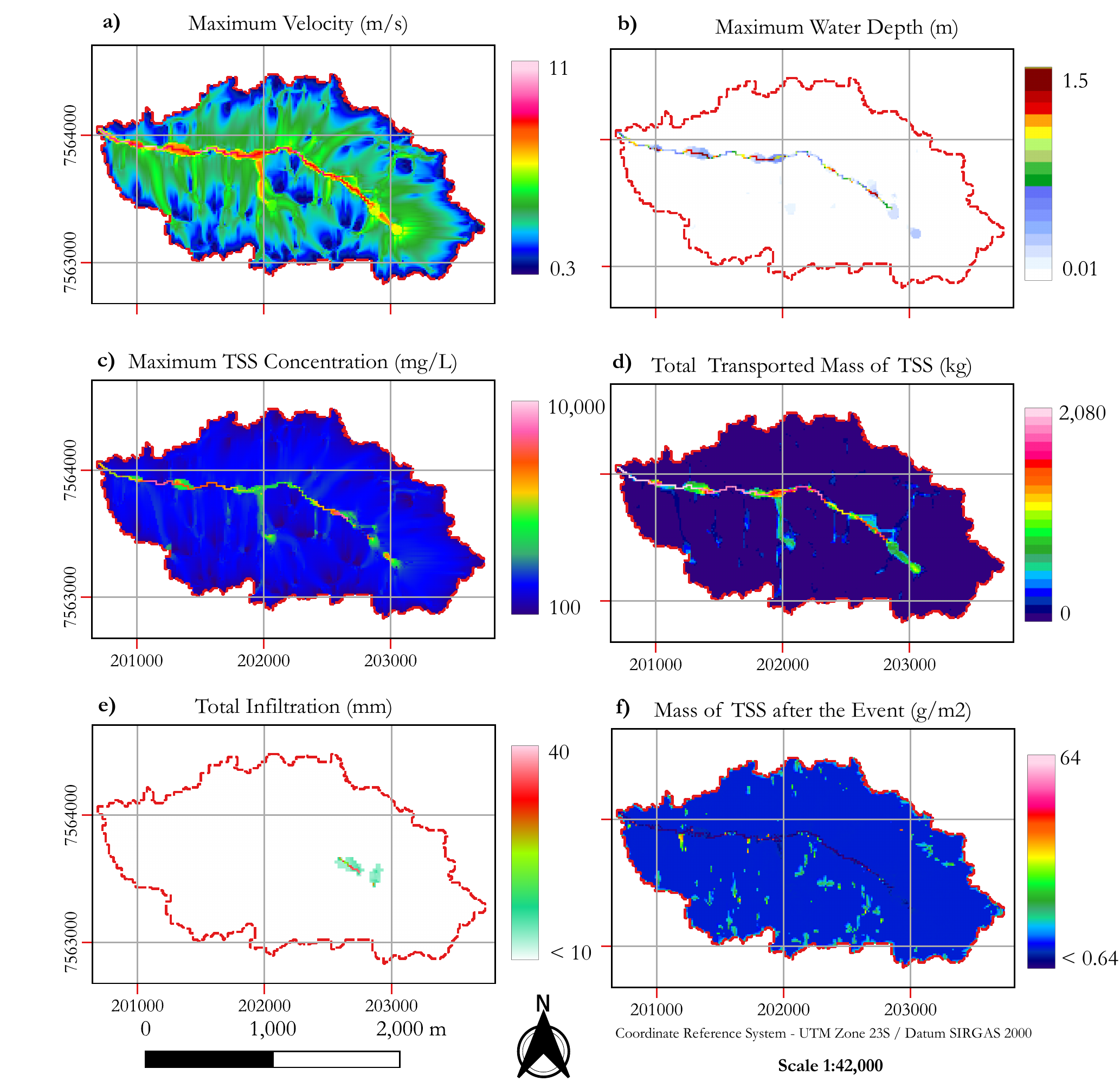}
    \caption{Simulation results with baseline scenario parameters, where a) is the maximum velocity, b) is the map of maximum depths, c) the maximum instantaneous concentration of TSS, d) is the map that represents the total mass that passed through each cell. The catchment boundaries is given by the red dashed lines.}
    \label{fig:baseline}
\end{figure*}

The simulation of the TPC for $1$ year return period event for rainfall and for the number of dry days is shown in Fig.~\ref{fig:baseline}. Fig.~\ref{fig:baseline} b) shows the maximum flood depth in the catchment, identifying areas susceptible to flooding with maximum depths of up to $1.50$ m for a $1$ hour of rainfall and $32$ mm of volume distributed in alternating blocks. Fig.~\ref{fig:baseline} a) shows the maximum velocity map, which exceeded $\mathrm{10~m/s}$ in the stream. Note that the maximum velocities are not necessarily associated with this maximum depth due to the rise and recession of the hydrographs with the propagation of the diffusive wave. The surface runoff generated was approximately equal to the total rainfall volume of $32$ mm, except for the volume infiltrated in permeable areas, illustrated in Fig.~\ref{fig:baseline} e). In this figure, it is possible to observe infiltrated volumes greater than the precipitated volume. This occurred because the pervious areas receive runoff volume from several cells upstream, which increases the ponding depth and therefore increases the infiltration capacity. Although most of the catchment is impervious, flood depths occurred mainly in the stream, falling toward the overbanks only in a few areas, as illustrated in Fig.~\ref{fig:baseline} b). This occurred due to the relatively low return period assumed in the modeling.

Regarding the TSS transport, Fig.~\ref{fig:baseline} d) shows how much pollutants have flowed to each cell during the event studied. Naturally, the stream is the area with the greatest passage of pollutants. However, it is possible to identify locations outside the urban stream that also have a high level of pollution transport. These results could be strategically used to identify possible candidate areas for the implementation of LIDs. Therefore, this methodology makes it possible to quantitatively identify the most suitable areas to maximize the capture of pollutants carried by surface runoff, especially the TSS. 

After the rainfall event, the remaining mass in the catchment is shown in Fig.~\ref{fig:baseline} f). This map illustrates the relatively clean stream and some areas with a relatively large accumulation of pollutants (e.g., $> 60$ g/m\textsuperscript{2} or $9.3$ kg of $\mathrm{TSS}$ in each pixel of $\mathrm{156.25~m^2}$). Therefore, this map can help identify areas of accumulation and can serve as information for model calibration when used for sediment modeling. Despite being more dynamic and instantaneous, the maximum concentration also allows one to identify the maximum polluting potential of surface runoff water, as illustrated in Fig.~\ref{fig:baseline} c).

\begin{figure*}
    \centering
    \includegraphics[scale = 0.30]{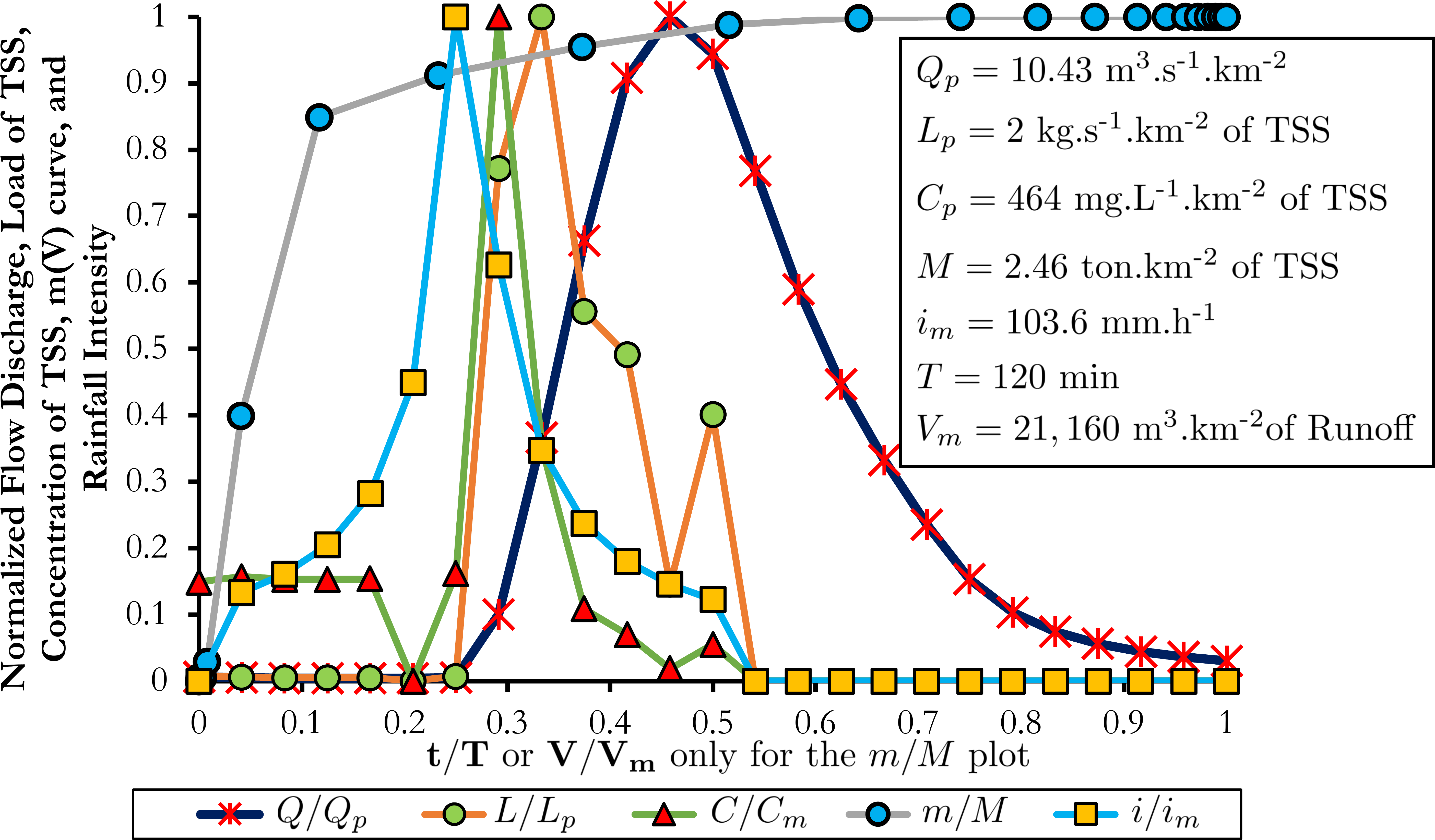}
    \caption{Normalized modeling results, where all values are divided by their next maximum values and are typically divided by the watershed area of $3.20$ km\textsuperscript{2}. \textit{Q} is the flow discharge, \textit{Q}\textsubscript{p} is the peak flow, \textit{L} is the pollutant load, \textit{L}\textsubscript{p} is the maximum pollutant load, \textit{C} is the pollutant concentration, \textit{C}\textsubscript{p} is the maximum pollutant concentration \textit{m} is the washed pollutant mass, \textit{M} is the total mass of pollutant washed, \textit{i} is the rainfall intensity, \textit{i}\textsubscript{m} is the maximum rainfall intensity, \textit{t} is the time and T is the total duration.}
    \label{fig:normalized}
\end{figure*}

The analysis of the normalized outlet hydrograph result is presented in Fig.~\ref{fig:normalized}. It is possible to observe the hysteresis phenomenon \citep{aich2014quantification}, which shows that the concentration peak occurs approximately $25$ minutes before the surface runoff peak. First, the rainfall peak occurs, following the concentration, load, and discharge peak, respectively. The first flush chart also shows that more than $90$\% of the TSS are washed in $30$\% of the volume. The same chart allows us to estimate (i) the time of concentration for this event, (ii) the peak time of flow discharges, concentrations, and loads, and allows comparison of results with other catchments, since all values are normalized by the catchment area.

\subsection{Challenges and Limitations of the Application of Distributed Models in Poorly-Gauged Catchments}
Depending on the purpose and scale of the study, elevation data may be crucial in applying hydrological and water quality models. In the case of modeling focused on the delineation of flood inundation maps, FEMA, the Federal Emergency Agency of the United States, recommends as a minimum criterion hydrodynamic simulations with a resolution of up to $3$ m with a vertical resolution of at most $1$ cm. Detailed elevation data are available in countries such as Brazil only in some large cities, e.g. São Paulo, making it difficult to apply them at several important points where floods occur \citep{santos2016suscetibilidade}.

The most recurring application of hydrodynamic models is the study and delineation of flood inundation areas \citep{do2021assessing, erena2018flood, fava2022linking}. Although the HydroPol2D model does not solve the complete Saint-Venant $2$D equations, its diffusive wave methodology is promising for determining flood areas in catchments where convective and local acceleration phenomena do not act as the main hydrodynamic governing processes. The flood inundation depth coupled with the velocity maps can serve as a basis to calculate the risk of human instability during a flood event \citep{rotava2013simulaccao}, to assess potential flood damage \citep{jamali2018rapid} or as input data for estimating the value of flood insurance policies \citep{aerts2011climate}. Furthermore, the model can be used to estimate the time of concentration without requiring calculating it by empirical formulae \citep{manoj2012estimating}, as previously presented. Additionally, flows at the catchment outlet can be estimated without the need for unit hydrographs.

Examples of the use of distributed models to determine hydrographs are presented in \cite{furl2018assessment, sharif2010application} and \cite{ sharif2013physically}. To this end, however, if the information on where the stream passes is dissolved in the coarse resolution of the elevation pixels, it is necessary to recondition the terrain model, smoothing thalweg lines and elevation peaks, or sometimes imposing lower elevations in channel sections as presented in this paper. Another application is the spatial assessment of infiltration, which can be important in some urban areas and plays a major role in rural areas. This analysis can aid in spatial quantification of infiltration, which can aid in the decision about the implanted or chosen crop \citep{paudel2011comparing}. These examples show that, although modeling aimed at delineating flooded areas via $2$D modeling requires high spatial resolution DEM, the determination of flows and, at least, the identification of critical points in the catchment can be identified with free data derived from satellite products (e.g., SRTM \citep{drusch2012sentinel} and Alos Palsar \citep{rosenqvist2007alos}).

An important point for water quality purposes is the lack of adequate time-scale water quality data to calibrate and validate high-resolution distributed models. For example, the complete calibration of the HydroPol2D model would require high-resolution information on rainfall (i.e., sub-daily), flow, and pollutants concentration \citep{gomes2021spatial}. The main information is the observed pollutograph, the rainfall hyetograph, and the outlet hydrograph. However, other information, such as initial soil moisture conditions and initial water depths, or internal boundary conditions, may be required in more complex cases.

The most uncertain variable and that is very difficult to estimate is the initial build-up map \citep{wijesiri2015process}. Several studies indicate that the use of the build-up equation with ADD as a dependent variable may not correctly represent the pollutant accumulation process in urban catchments \citep{bonhomme2017should, zhang2019testing}. Variables such as the predominant wind speed and direction, atmospheric pressure, humidity, and the geographic position of the catchment near roads and highways, among others, can play an essential role in the accumulation of pollutants \citep{pandey2016physically}. Furthermore, the build-up model assumes a uniform accumulation for each land use, disregarding accumulation characteristics (e.g., source pollution release). All these limitations must be taken into account when modeling water quality. The HydroPol2D model, although developed for non-point source pollution, allows the modeling of source pollution by entering load rates at specific cells as external boundary conditions.

Despite the difficulties in model calibration, most parameters can be estimated, at least at the preliminary analysis level, based on the literature \citep{rossman2010storm}. Sensitivity analysis reveals that the most important parameters of the model are the Manning's roughness coefficients and the wash-off coefficients, especially the exponent (C\textsubscript{4}). Both parameters can be derived as a function of land use classifications. The model makes it possible to identify, using mostly physically-based equations, the hydrological, hydrodynamic, and distribution behavior of diffusive pollution in catchments where Hortonian processes govern the flow. The model allows for the estimation of important factors at the outlet level and spatialized values throughout the catchment. Therefore, one of the applications is to determine the critical areas of accumulation of pollutants in the catchment during and after precipitation events. This information can be used in master plans for better water quality management and to define potential areas to implement LID techniques focused on treating part of surface runoff \citep{batalini2022low, mcclymont2020towards, de2021different}. 

Furthermore, the model can be used to evaluate the spatial impact of LIDs at the watershed scale by modeling its pixels with different land use and elevation properties (e.g., reducing the pixel elevation to simulate the ponding layer on the surface). This analysis can be done to quantify water quality and estimate the volumes of surface runoff retention. Furthermore, at the outlet of the catchment, dynamic factors such as the load and concentration of pollutants are estimated and are indicators of the response of the catchment to simulated events. Finally, the first flush modeling can be performed using the HydroPol2D model, which is an important evaluation for urbanized catchments.

\section{Conclusions} \label{sec:conclusions}
Evaluating the impacts of surface runoff quality and quantity in urbanized catchments requires the temporal and spatial quantification of flood depths, pollutant transport, and fate. With this focus, the HydroPol2D model was designed and first applied in the V-tilted catchment to identify the role of the maximum flow velocity limitation. Our results indicates that limiting velocities to critical, reduce the model performance. The HydroPol2D water quality module was calibrated and validated with the observed data provided from a wooden board catchment. Subsequently, the model was applied in the Tijuco Preto catchment in São Carlos - focusing on the qualitative and quantitative quantification of the spatial-temporal behavior of surface runoff. Even with the lack of observed or high-resolution elevation data, it was possible to evaluate the quali-quantitative dynamics of the stormwater runoff for a return period of 1 year, both for rainfall and the number of antecedent dry days. An event composed of drought followed by a flood was evaluated. The results of the numerical simulation for the Numerical Case Study 3 indicate the following:

\begin{itemize}
    \item The maximum load and the maximum TSS concentration at the outlet are $8.22 \pm 1.29$ kg.s\textsuperscript{-1} and $1,460 \pm 832$ mg.L\textsuperscript{-1}. Normalizing by the catchment area of $3.20$ km\textsuperscript{2} it follows that the maximum concentration of TSS is $456 \pm 260$ mg.L\textsuperscript{-1}.km\textsuperscript{-2} and the maximum load of TSS is $2.56 \pm 0.4$ kg.s \textsuperscript{-1}.km\textsuperscript{-2} for a $1$-yr flood-drought event.
    \item The Washoff-Ratio coefficient and the $\mathrm{EMC}$ were $0.63 \pm 0.11$ and $131.59 \pm 16$ mg.L\textsuperscript{-1}, respectively, for a $1$-yr flood-drought event.
    \item The volume of TSS washed in $30$\% of the runoff volume was $89\% \pm 10$\%, indicating a high first-flush phoenomen in the catchment, considering an uncerntainty in wahs-off parameters from $-40$\%  to $40$\%.
\end{itemize}
The results of this article show how quali-quantitative modeling can be used to determine possible areas for applying LIDs, delineating areas prone to flooding, analysis of maximum flow velocities, and therefore risk of human instability due to floods. Furthermore, it allows to identify maps of maximum pollutant concentration. Despite the impossibility of calibrating the model for the TPC catchment due to lack of data, the calibration of quali-quantitative parameters is encouraged and can be done in the model via automatic calibration using optimization packages in Matlab ® \citep{higham2016matlab}. Furthermore, the analysis performed can be replicated for other combinations of RPs for rainfall and antecedent dry days. Future studies will incorporate resilience metrics to floods and water quality to aid decision-making in warning systems. Moreover, future work will incorporate modeling via continuous simulation with spatially varied precipitation. Finally, testing the simulation time performance of the model against state-of-the-art software is also desired.

As in other distributed models, the challenge for the quality of the results presented by HydroPol2D is related to the quality of the input data, especially the topography and land use and land cover data. However, the model requires relatively few parameters to describe the hydraulic properties of the terrain and allows us to simulate quantity and/or quality. When simulating only water quantity, significant differences in processing time are obtained with HydroPol2D. Another advantage is the model's applicability, which, if the hydrological processes are predominantly Hortonian, allows simulating catchments at all spatial scales. Future studies will incorporate spatial variability of rainfall and evapotranspiration for large-scale watersheds. Ultimately, the HydroPol2D model can become a tool for real-time forecasting by incorporating distributed modeling of hydrodynamics and pollutant transport and fate. 

\section*{Acknowledgment}
The authors appreciate the support of the City of San Antonio, by the San Antonio River Authority, CAPES Ph.D Scholarship, and the PPGSHS PROEX Graduate Program.

\section*{Appendix A. Supplementary Material}
Supplementary data related to this article can be found at \url{https://github.com/marcusnobrega-eng/HydroPol2D}. \

\section*{Data Availability Statement}
Algorithms and data used are available in an open repository at \cite{HydroPol}.




\printcredits

\bibliographystyle{cas-model2-names}
\bibliography{cas-refs.bib}



\end{document}